\documentclass[journal]{IEEEtai}

\usepackage[colorlinks,urlcolor=blue,linkcolor=blue,citecolor=blue]{hyperref}

\usepackage{color,array}

\usepackage{graphicx}
\usepackage{mathtools}
\usepackage{amsmath}
\usepackage{amssymb}
\usepackage{algorithm,algorithmic}
\usepackage{cite}

\begin{document}

\title{Application of Deep Reinforcement Learning\\to Event-Triggered Control\\ for Networked Artificial Pancreas Systems}

\author{Junya  Ikemoto, Satoshi Maruyama, and Kazumune Hashimoto, \IEEEmembership{Member, IEEE}

\thanks{This work was partially supported by JST ACT-X JPMJAX23CK, JST BOOST JPMJBY25E0, and JSPS KAKENHI Grant Numbers 26K21173, 25K07794, 22KK0155. }

\thanks{J. Ikemoto is with Division of Electrical, Electronic and Infocommunications Engineering, Graduate School of Engineering, University of Osaka, Suita Yamadaoka 2-1, Japan (e-mail: ikemoto@eei.eng.osaka-u.ac.jp).}
\thanks{The authors acknowledge the use of ChatGPT (OpenAI) for language refinement and editorial assistance in preparing this manuscript, and for generating a portion of the illustration in Fig. 3. All content was reviewed, revised, and verified by the authors.}

}


\maketitle

\begin{abstract}
This paper proposes a deep reinforcement learning (DRL)-based event-triggered controller design for networked artificial pancreas (AP) systems. In such systems, it is necessary to handle nonlinear glucose–insulin dynamics under limited computational resources while carefully managing the communication frequency to ensure energy-efficient operation. In recent years, DRL has attracted attention as a promising approach for controller design in AP systems. However, most existing DRL-based approaches assume periodic control updates without considering communication efficiency. Effective management of communication frequency is essential for energy-efficient operation. In this context, hierarchical event-triggered DRL methods have been proposed to jointly learn policies for control inputs and communication scheduling. However, in AP systems with limited observations, it is challenging to jointly learn both policies through trial and error. To address this challenge, we propose a practical DRL-based event-triggered controller design that avoids explicitly learning communication scheduling by introducing a rule-based triggering criterion based on changes in blood glucose. As a result, decision-making occurs at irregular intervals, and the problem is formulated as a semi-Markov decision process (SMDP), for which we extend a standard DRL algorithm. Numerical experiments demonstrate that the proposed method reduces communication frequency while maintaining control performance. More broadly, this study provides a new perspective on integrating DRL-based control with resource-aware design. This perspective is relevant to a wide range of systems that must operate under limited communication and energy resources.
\end{abstract}

\begin{IEEEImpStatement}
This study addresses an important but insufficiently explored challenge in AI-based artificial pancreas systems, namely the trade-off between control performance and communication efficiency in wireless operation. While artificial pancreas systems have enabled automated insulin delivery through wearable and wireless technologies without wired connections, their practical deployment can be constrained by energy limitations associated with frequent data transmission. By introducing an AI-based control framework that explicitly accounts for communication frequency, this study advances the design of more energy-efficient and reliable glucose control systems. This contribution can improve battery life and reduce the effort needed by patients by minimizing unnecessary device operation and data transmission, thereby making diabetes management easier to use in daily life. More broadly, the proposed approach highlights the importance of incorporating resource awareness into AI-based control systems, with potential applications in a wide range of wearable and networked technologies.
\end{IEEEImpStatement}


\section{Introduction}
\IEEEPARstart{D}{iabetes} mellitus (DM) is a metabolic disorder characterized by chronic hyperglycemia resulting from insufficient insulin secretion or impaired insulin action, which increases the risk of complications affecting multiple organs. DM is classified into type 1 diabetes (T1D) and type 2 diabetes (T2D) \cite{DM1,DM2}. T1D is a disease characterized by failure of the pancreatic $\beta$-cells, resulting in insufficient insulin production. T2D is primarily associated with insulin resistance. In T1D, exogenous insulin therapy, such as multiple daily injections and continuous subcutaneous insulin infusion, is essential for survival \cite{Care_DM}. Since blood glucose levels continuously fluctuate due to various factors, such as meals and physical activity, manually adjusting insulin dosing imposes a significant burden on patients.

To reduce the burden of blood glucose management for patients with diabetes, artificial pancreas (AP) systems have been actively developed. In an AP system, a continuous glucose monitor (CGM) \cite{CGM} continuously measures the patient’s blood glucose level, and a controller determines the appropriate insulin infusion rate based on these measurements and delivers insulin via an insulin pump to maintain blood glucose levels within the target range \cite{AP1,AP2}. Traditionally, to design controllers for AP systems, proportional–integral–derivative (PID) control \cite{PID_AP} has been applied. To facilitate the development of such systems and evaluation in silico, the UVA/Padova T1D Simulator has been developed based on mathematical models that describe physiological glucose–insulin dynamics \cite{UVA_1,UVA_2}. Using the high-fidelity of such simulators, model-based controller design methods, including model predictive control (MPC), have been actively studied \cite{MPC_AP1,MPC_AP2,MPC_AP3,MPC_AP4,MPC_AP5}.

\begin{figure}[htbp]
\centering
\includegraphics[width=8.5cm]{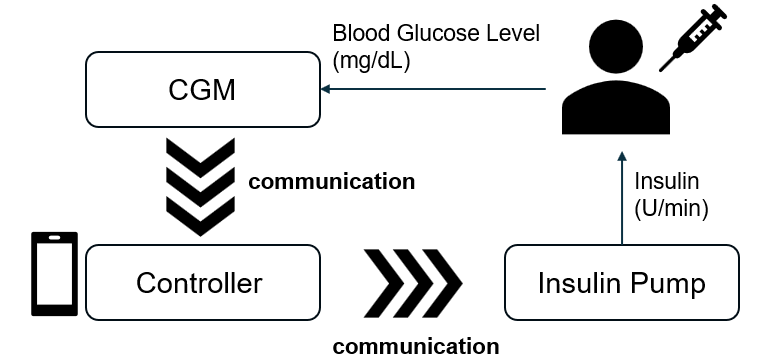}
\caption{Illustration of a networked AP system. The controller, implemented on a wearable device, receives measurements from the CGM, updates the insulin infusion rate, and delivers insulin via an insulin pump.}
\label{fig:ap_system}
\end{figure}

In recent years, as illustrated in Fig. \ref{fig:ap_system}, networked AP systems, in which a CGM, a controller, and an insulin pump exchange information over a network, have been actively studied \cite{NCS_AP1,NCS_AP2}. By reducing the need for wired connections, these systems improve user comfort and mobility. In addition, they enable the use of remote computational resources to implement computationally intensive control algorithms such as robust and nonlinear MPC. Furthermore, the use of cloud-based resources facilitates the efficient and flexible updating of models and control algorithms. However, the use of overly computationally intensive methods increases the power consumption of the controller, which can lead to heat generation and frequent charging. Consequently, controllers for AP systems are often designed using linear MPC. Since glucose–insulin dynamics are inherently nonlinear and subject to unpredictable disturbances, such as meals and physical activity, advanced control algorithms are increasingly required. Ideally, such algorithms should achieve both high control performance and low real-time computational cost.

\begin{figure}[htbp]
\centering
\includegraphics[width=8.5cm]{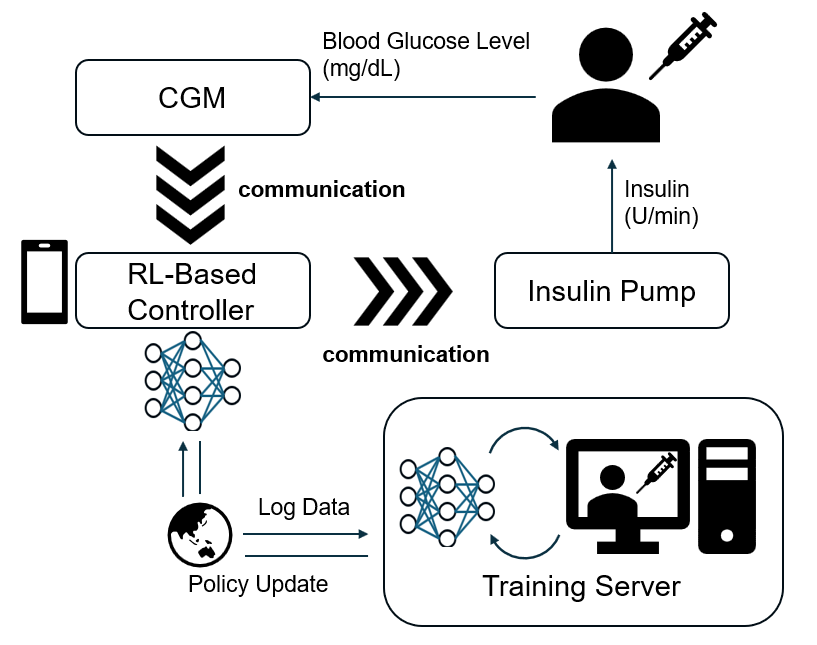}
\caption{An example architecture of a networked AP system with an RL-based controller. The policy is trained on a remote server using a simulator, such as the UVA/Padova T1D simulator. The trained policy is then deployed on a wearable device for automated insulin dosing.}
\label{fig:drl_ap_system}
\end{figure}

As an alternative to methods involving computationally intensive online optimization, reinforcement learning (RL) has recently emerged as a promising approach for designing advanced controllers \cite{MPC_and_RL}. RL is a machine learning framework for sequential decision-making problems \cite{RL_book}. In RL, a learner, which is called an {\it agent}, interacts with an {\it environment} (or a system) and learns a policy automatically. RL-based controller design requires substantial computational resources during the training phase. However, once training is completed, control actions can be computed without online optimization procedures. Moreover, deep RL (DRL), which uses deep neural networks (DNNs) as function approximators, can design controllers for high-dimensional nonlinear systems \cite{DRL_book}. However, due to safety concerns and cost constraints, RL is typically conducted in simulation, which introduces a gap between simulated and real environments. To address this issue, many techniques have been proposed to design controllers robust to various simulation gaps, including {\it domain randomization} (DR), which improves robustness by randomizing simulation environments during training \cite{DR1,DR2}. This has facilitated the application of DRL to medical systems, where direct trial-and-error on real patients is difficult. An RL-based networked AP system has been proposed in \cite{Lee}, demonstrating a framework in which the controller is trained and evaluated in a simulation environment, with potential for real-world deployment, as illustrated in Fig. \ref{fig:drl_ap_system}.

Networked AP systems can be regarded as a class of networked control systems (NCSs), which have been extensively studied in the control systems community \cite{NCSs}. In NCSs, in addition to achieving control objectives, managing communication frequency is important to reduce excessive power consumption. For networked AP systems, reducing communication frequency is also important for wearable devices with limited battery capacity. Event-triggered control (ETC) is a reactive strategy that reduces communication frequency by updating control inputs only when a predefined triggering condition is satisfied \cite{etc_stc}. 
In other domains such as robotics, DRL-based methods incorporating ETC have also been proposed \cite{ETC_DRL_1,ETC_DRL_2,ETC_DRL_3}. However, directly applying these methods to networked AP systems is challenging, as it requires jointly learning both insulin dosing and event-triggering decisions, which significantly increases the difficulty of the learning problem, particularly under limited observability, where only CGM measurements are available. To address this challenge, we introduce a rule-based event-triggering mechanism defined by the magnitude of changes in blood glucose levels, where an event is triggered when the change exceeds a fixed or variable threshold. This decouples the event-triggering mechanism from the agent's decision-making, thereby alleviating the complexity of the learning problem. As a result, decision intervals become irregular, which cannot be adequately handled within the standard {\it Markov decision process} (MDP) framework. Therefore, we formulate the problem as a {\it semi-Markov decision process} (SMDP) \cite{SMDP} and extend a DRL algorithm accordingly.

\subsection{Contributions}
Our main contributions are summarized as follows:
\begin{itemize}
    \item We develop a DRL-based event-triggered controller design for energy-efficient networked AP systems.
    \item We propose a practical DRL-based controller design that eliminates the need to explicitly learn the event-triggering mechanism by introducing a rule-based criterion, thereby reducing the complexity of the learning problem.
    \item We extend a standard DRL algorithm to handle problems formulated as SMDPs.
    \item Through numerical simulations, we demonstrate improved communication efficiency while maintaining control performance and analyze the impact of thresholds on control performance.
    \item We show that an appropriate event-triggered mechanism can facilitate policy learning in AP systems.
\end{itemize}
Existing DRL-based studies on AP systems have primarily assumed periodic control updates without considering communication frequency. To the best of our knowledge, this study is among the first to develop a practical DRL-based event-triggered controller design for energy-efficient networked AP systems.

\subsection{Related Works}
\subsubsection{Controller Design for Artificial Pancreas Systems}
To reduce the burden of insulin administration for patients with diabetes, feedback controllers for AP systems have been developed. PID control, which is widely used in industrial plants, is a standard approach and its clinical trials were conducted in the 2000s \cite{PID_AP}. However, since glucose–insulin systems are nonlinear and sensitive to disturbances, tuning the PID gains is generally challenging. Fuzzy logic–based controller designs have also been proposed as a class of soft computing approaches \cite{fuzzy_AP}. However, their performance depends on the choice of fuzzy rules and membership functions. With the development of high-fidelity mathematical models, such as the UVA/Padova simulator \cite{UVA_1,UVA_2}, MPC-based methods have become a widely adopted approach \cite{MPC_AP1,MPC_AP2,MPC_AP3,MPC_AP4,MPC_AP5}. However, in practice, due to real-time computational constraints, controller design has been largely limited to linear MPC. To address these limitations, RL has increasingly been explored as a control design approach for handling nonlinearities, model uncertainties, and disturbances with reduced real-time computational requirements \cite{Lee,RL_AP1,RL_AP2,RL_AP3,RL_AP4, RL_AP5}. Specifically, in \cite{Lee}, Lee et al. propose a DRL-based controller design architecture implemented as a configuration of AndroidAPS \cite{Android_APS}, which is a smartphone-hosted Do-It-Yourself AP system. However, existing DRL-based studies on AP systems have primarily assumed periodic control updates, without explicitly addressing communication efficiency in networked implementations.

\subsubsection{Deep Reinforcement Learning for Energy-efficient Networked Control}
DRL-based methods for the design of networked controllers have been proposed to improve communication efficiency. In \cite{STC_DRL_1,STC_DRL_2}, DRL is applied to {\it self-triggered control} (STC), in which the policy simultaneously determines both the control input and the next update time at each decision step. However, STC is limited in handling unpredictable disturbances. Consequently, it is challenging to apply STC to AP systems, where disturbances such as meals and physical activity occur frequently. In contrast, ETC can respond more effectively to such disturbances. Thus, in this study, we focus on an ETC-based approach for AP systems. Specifically, in \cite{ETC_DRL_1,ETC_DRL_2,ETC_DRL_3}, hierarchical DRL algorithms for ETC are proposed, where the upper layer determines whether an update of the control input is necessary and the lower layer determines the control input only when required. However, simultaneous learning of both policies increases the complexity of the problem, particularly in AP systems with limited observability, where learning relies solely on CGM measurements. In contrast to these approaches, our method avoids jointly learning the triggering policy and instead relies on a rule-based mechanism, which significantly simplifies the learning problem. 

\subsection{Structure}
The remainder of this paper is organized as follows. Section II reviews RL, DRL, and DRL-based networked controller design. Section III describes AP systems and formulates the control problem. Section IV presents a DRL-based controller design method for networked AP systems. Section V demonstrates the effectiveness of the proposed method through numerical simulations. Section VI concludes the paper and discusses future research directions.

\subsection{Notation}
$\mathbb{N}$ denotes the set of natural numbers. $\mathbb{R}$ denotes the set of real numbers. $\mathbb{R}_{\ge0}$ denotes the set of non-negative real numbers. $\mathbb{R}^{n}$ denotes the $n$-dimensional Euclidean space. $|S|$ denotes the cardinality of a set $S$.

\section{Preliminaries}

\subsection{Reinforcement Learning}
RL is a machine learning approach for sequential decision-making problems\cite{RL_book}. In RL, an agent learns a policy to determine actions through interaction with an environment. A decision-making problem is formulated by an MDP: $\mathcal{M}=\left<\mathcal{S},\mathcal{A},R,p_f,\rho_{0}\right>$, where $\mathcal{S}$ is the state space of an environment, $\mathcal{A}$ is the action space of an agent, $R:\mathcal{S}\times\mathcal{A}\to\mathbb{R}$ is the immediate reward function, $p_f:\mathcal{S}\times\mathcal{A}\times\mathcal{S}\to\mathbb{R}_{\ge0}$ is the probability density of state transitions, and $\rho_0:\mathcal{S}\to\mathbb{R}_{\ge0}$ is the initial state distribution. In RL, a deterministic policy $\mu:\mathcal{S}\to\mathcal{A}$ or a stochastic policy $\pi:\mathcal{S}\times\mathcal{A}\to\mathbb{R}_{\ge0}$ is learned to maximize the following discounted return $G_{h}=\sum_{k=0}^{\infty}\gamma^{k}R(s_{h+k},a_{h+k})$, where $\gamma\in(0,1)$ is a discount factor, $s_h\in\mathcal{S}$ and $a_h\in\mathcal{A}$ are the state and action at time $h\in\mathbb{N}$, respectively. To handle stochasticity, we commonly use the following functions.
\begin{eqnarray}
    V^{\pi}(s)&=&E_{\tau_h\sim\rho_{\pi}}\left[G_h|\ s_h=s\right],\ \forall s\in\mathcal{S},\label{value_function}\\
    Q^{\pi}(s,a)&=&E_{\tau_h\sim\rho_{\pi}}\left[G_h|\ s_h=s,a_h=a\right],\nonumber\\
    &&\hspace{3cm}\forall (s,a)\in\mathcal{S}\times\mathcal{A}.\label{Q_function}
\end{eqnarray}
$V^{\pi}$ and $Q^{\pi}$ are called the state value function underlying $\pi$ and the state-action value function (Q-function) underlying $\pi$, respectively. $\tau_h$ denotes a trajectory $s_h,a_h,r_h,s_{h+1},...$ and $\rho_{\pi}$ is the following joint probability density function.
\begin{eqnarray}
    \rho_{\pi}(\tau_h)=\prod_{t=h}^{\infty}\pi(a_t|s_t)p_{f}(s_{t+1}|s_{t},a_{t}).\nonumber
\end{eqnarray}
Additionally, an advantage function underlying $\pi$ is defined as follows:
\begin{eqnarray}
    A^{\pi}(s,a)=Q^{\pi}(s,a)-V^{\pi}(s).\label{advantage_function}
\end{eqnarray}
From an intuitive perspective, the advantage function measures how much better a specific action $a\in\mathcal{A}$ is compared to the expected performance of the policy $\pi$. The state-value function and the Q-function are known to satisfy the following recursive equations, which are called {\it Bellman equations}.
\begin{eqnarray}
    V^{\pi}(s)&=&E_{a\sim \pi(\cdot|s)}\left[R(s,a)+\gamma E_{s'\sim p_{f}(\cdot|s,a)}\left[V^{\pi}(s')\right]\right],\nonumber\\
    \label{Bellman_eq_V}\\
    Q^{\pi}(s,a)&=&R(s,a)+\gamma E_{s'\sim p_{f}(\cdot|s,a),a'\sim \pi(\cdot|s')}\left[Q^{\pi}(s',a')\right].\nonumber\\
    \label{Bellman_eq_Q}
\end{eqnarray}
The objective of RL can be formulated as follows:
\begin{eqnarray}
\pi^{*}\in\arg\max_{\pi}\ E_{s_0\sim\rho_0}\left[V^{\pi}(s_0)\right].\label{RL_goal}
\end{eqnarray}
In the case where $\mathcal{S}$ and $\mathcal{A}$ are continuous state and action spaces, value functions and policies are typically approximated using parametric function approximators, such as DNNs, whose parameters are optimized. For example, in the {\it REINFORCE} algorithm \cite{RL_book}, the policy is parameterized and the following objective is maximized.
\begin{eqnarray}
    J(\theta)=E_{s_0\sim\rho_0}\left[V^{\pi_{\theta}}(s_0)\right],\label{REINFORCE_objective}
\end{eqnarray}
where $\theta$ is a parameter vector of the policy $\pi_{\theta}$. To update $\theta$ using a gradient-based approach, we need to obtain the gradient of $J(\theta)$ with respect to $\theta$. The gradient can be computed as
\begin{eqnarray}
    \nabla_{\theta} J(\theta)=E_{\tau\sim\rho_\pi}\left[\sum_{h=0}^{\infty}\nabla_{\theta}\log \pi_{\theta}(a_h|s_h)Q^{\pi_{\theta}}(s_h,a_h)\right],\label{PG}
\end{eqnarray}
which is called {\it policy gradient} \cite{PG}. Since the true Q-value is unavailable during training, it is commonly approximated using {\it Monte Carlo estimates} of the return obtained from interactions with the environment. However, Monte Carlo estimates suffer from high variance, which can make the learning process unstable. To reduce variance, the {\it Actor–Critic} approach is commonly employed. In this approach, the value function $V^{\pi_{\theta}}$ is approximated by a parametric model $V_{\phi}$, and $\pi_{\theta}$ and $V_{\phi}$ are updated simultaneously. Here, $\phi$ denotes a parameter vector of the approximated value function. We use the following property: for an arbitrary function $b:\mathcal{S}\to\mathbb{R}$, 
\begin{eqnarray}
    E_{a_h\sim\pi_{\theta}}\left[\nabla_{\theta}\log \pi_{\theta}(a_h|s_h)b(s_h)\right]=0.\label{baseline}
\end{eqnarray}
Using this property, an arbitrary baseline $b(s_h)$ can be subtracted from the Q-function without introducing bias. Thus, the policy gradient can be written as
\begin{eqnarray}
&&\nabla_\theta J(\theta)\nonumber\\
&=& E_{\tau \sim \rho^{\pi_\theta}}
\left[
\sum_{h=0}^{\infty}
\nabla_\theta \log \pi_\theta(a_h|s_h)
\left(Q^{\pi_\theta}(s_h, a_h) - b(s_h)\right)
\right].\nonumber
\end{eqnarray}
In particular, it is known that choosing $b(s_h)$ close to the value function $V^{\pi_{\theta}}(s_h)$ reduces variance. Thus, the policy gradient can be expressed as
\begin{eqnarray}
\nabla_\theta J(\theta)=E_{\tau \sim \rho^{\pi_\theta}}
\left[\sum_{h=0}^{\infty}\nabla_\theta \log \pi_\theta(a_h|s_h)\,
A^{\pi_\theta}(s_h,a_h)\right].
\end{eqnarray}
$A^{\pi_\theta}(s_h,a_h)$ is not directly available. Therefore, it is estimated using $V_{\phi}$ and sampled trajectories. For example, we employ the {\it generalized advantage estimator} (GAE) to compute the advantage \cite{GAE}, defined as
\begin{eqnarray}
\hat{A}_{\phi}(s_h,a_h)
&=& \sum_{l=0}^{\infty} (\gamma \lambda)^l \, \delta_{h+l}, \\
\delta_h &=& r_h + \gamma V_{\phi}(s_{h+1}) - V_{\phi}(s_h),
\end{eqnarray}
where $\delta_h$ denotes the {\it temporal-difference error} and $\lambda \in [0,1]$ is a parameter that controls the bias–variance trade-off. When $\lambda=0$, it reduces to {\it TD-learning}, whereas $\lambda=1$ corresponds to the Monte Carlo method. 

In practice, given a finite trajectory $(s_0, a_0, r_0, \ldots, s_N)$ generated by the current policy $\pi_{\theta}$, the advantage values can be efficiently computed by performing the following backward recursion
\begin{eqnarray}
    \hat{A}_{h}&=&\delta_{h}+\gamma\lambda\hat{A}_{h+1}\nonumber,\\
    \hat{A}_{N}&=&0.\nonumber
\end{eqnarray}
The value function $V_{\phi}$ is trained by minimizing the mean squared error between $V_{\phi}(s_h)$ and the empirical return $\hat{G}_h=\sum_{k=h}^{N}\gamma^{k-h}r_{k}$, which is often approximated as $\hat{G}_h \simeq V_{\phi_{\text{old}}}(s_h)+\hat{A}_{h}$, where $V_{\phi_{\text{old}}}$ denotes the value function before the update and is treated as fixed when constructing the target.

\subsection{Deep Reinforcement Learning}
DRL refers to RL with DNNs used as function approximators. By leveraging the powerful function approximation capability of DNNs, complex decision-making problems with high-dimensional state and action spaces can be addressed. For problems with continuous state and action spaces, various DRL algorithms with the Actor–Critic approach have been proposed \cite{A3C,TRPO,PPO}. The {\it trust region policy optimization} (TRPO) algorithm constrains the step size of policy updates using {\it Kullback–Leibler} (KL) divergence to ensure monotonic policy improvement \cite{TRPO}. Although the TRPO algorithm is effective, it requires computationally intensive procedures, such as the conjugate gradient method and line search, which limit the direct application of standard optimization methods, such as {\it stochastic gradient descent} (SGD) or {\it Adam} \cite{Adam}. The {\it proximal policy optimization} (PPO) algorithm is a simplified variant of the TRPO algorithm that allows for stable policy updates while reducing computational complexity \cite{PPO}. In the PPO algorithm, the probability ratio is defined as $\rho_h(\theta) = \frac{\pi_{\theta}(a_h|s_h)}{\pi_{\theta_{\text{old}}}(a_h|s_h)}$. Instead of solving a constrained optimization problem, the probability ratio is clipped. The policy $\pi_{\theta}$ is updated to maximize the following objective using the trajectories $(s_0,a_0,r_0,s_1,\ldots,s_N)$ generated by the policy $\pi_{\theta_{\text{old}}}$.
\begin{eqnarray}
    J(\theta)=L_{\text{clip}}(\theta)+c_{\text{ent}}L_{\text{ent}}(\theta), \label{PPO_actor}
\end{eqnarray}
where
\begin{eqnarray}
&&L_{\text{clip}}(\theta)=
\frac{1}{N}\sum_{h=0}^{N-1}\min\Big(\rho_h(\theta)A^{\pi_{\theta_{\text{old}}}}(s_h,a_h),\nonumber \\
&&\hspace{1cm}\text{clip}(\rho_h(\theta),\, 1-\epsilon,\, 1+\epsilon)A^{\pi_{\theta_{\text{old}}}}(s_h,a_h)
\Big),\label{clip_objective}
\end{eqnarray}
$L_{\text{ent}}(\theta)$ denotes the entropy bonus, and $c_{\text{ent}}>0$ is its coefficient. Since $A^{\pi_{\theta_{\text{old}}}}(s_h,a_h)$ is not directly available, it is estimated using GAE, as described in Section II.B. The value function is updated by minimizing the following mean squared error.
\begin{eqnarray}
    J(\phi)=\frac{1}{N}\sum_{h=0}^{N-1}\left(V_{\phi}(s_h)-\hat{G}_{h}\right)^{2},\label{PPO_critic}
\end{eqnarray}
where $\hat{G}_{h}$ denotes the empirical return or its approximation.

PPO is a widely used DRL algorithm, for which numerous implementation details and empirical findings have been reported \cite{PPO_1,PPO_2}. In this paper, we adopt the PPO algorithm as a baseline algorithm.

\subsection{DRL-based Networked Controller Design}
We consider the following continuous-time system.
\begin{eqnarray}
    \dot{x}(t)=f(x(t),u(t),d(t)),\label{dynamical_system}
\end{eqnarray}
where $x(t)\in\mathbb{R}^{n_x}$, $u(t)\in\mathbb{R}^{n_u}$, and $d(t)\in\mathbb{R}^{n_d}$ denote the state of the system, the control input, and the disturbance at time $t\in\mathbb{R}_{\ge0}$, respectively. $f:\mathbb{R}^{n_x}\times\mathbb{R}^{n_u}\times\mathbb{R}^{n_d}\to\mathbb{R}^{n_x}$ represents the dynamics of the system. The sensor provides measurements at each sampling period $\Delta>0$ as follows:
\begin{eqnarray}
y_h=g(x(h\Delta))+\epsilon_h,\ \ \ h\in\mathbb{N}\nonumber
\end{eqnarray}
where $g:\mathbb{R}^{n_x}\to\mathbb{R}^{n_y}$ is the output function that maps the state to the measurement, and $\epsilon_h$ denotes the measurement noise. The digital controller computes the control input $u_h$ based on the measurements $y_h$ and the control policy, and applies it to the system as follows:
\begin{eqnarray}
    u(t)=u_h,\ \ \ \forall t\in[h\Delta, (h+1)\Delta),\nonumber
\end{eqnarray}
which is referred to as a {\it zero-order hold} (ZOH). In general, unlike linear systems, the analytical design of digital controllers for nonlinear continuous-time systems is more challenging than that for linear systems. Instead, nonlinear MPC computes the control input by discretizing the continuous-time system using numerical integration, such as the {\it Runge–Kutta method} \cite{Numerical_ODE}, and solving a finite-horizon optimization problem online, which is generally nonconvex. However, this incurs a significant real-time computational cost due to the need for online nonconvex optimization \cite{MPC_and_RL}. As an alternative to such model-based approaches, DRL provides a data-driven framework for controller design based on trial-and-error interactions.

In addition, in NCSs, where the control loop is closed over a communication network, reducing communication frequency is important for improving energy efficiency. ETC is an effective approach for this purpose \cite{etc_stc}. Several DRL-based methods have been proposed for designing networked controllers with ETC \cite{ETC_DRL_1,ETC_DRL_2,ETC_DRL_3}. In these algorithms, the upper-level policy determines at $t=h\Delta$, whether to update the control input ($e_h=1$) or retain the previous control input ($e_h=0$). The lower-level policy determines a new control input $u_h$ only when the upper-level policy selects $e_h=1$. The control input applied to the system is given by
\begin{eqnarray}
    u(t)=\begin{cases}
        u_{h}, & e_h=1,\\
        \tilde{u}, & e_{h}=0,
    \end{cases}\ \ \ t\in[h\Delta,(h+1)\Delta),\label{etc_control_input}
\end{eqnarray}
where $\tilde{u}$ denotes the most recently applied control input.

\section{Controller Design for Artificial Pancreas Systems}
\subsection{Glucose-Insulin Dynamics}
\begin{figure}[htbp]
\centering
\includegraphics[width=8.5cm]{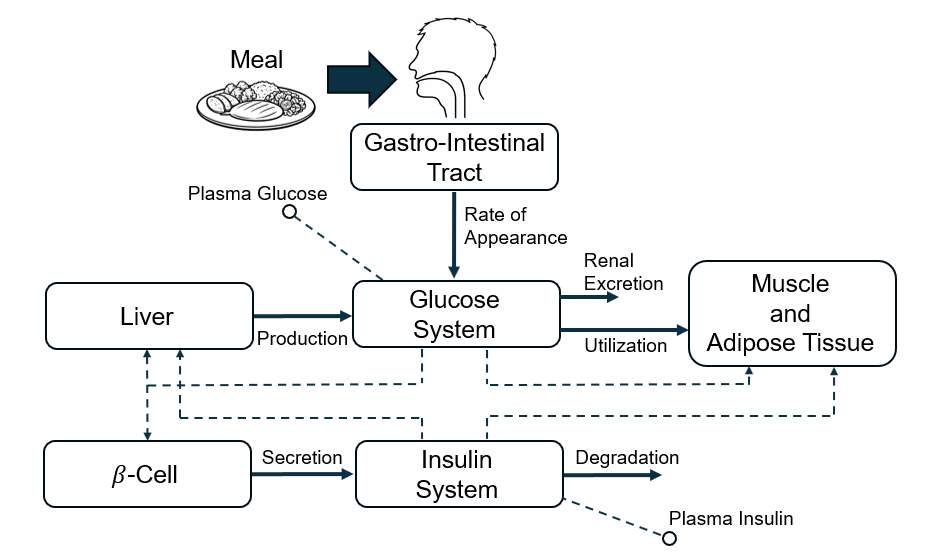}
\caption{Illustration of the glucose-insulin dynamics (S2008). The mathematical model represents the relationship among measured plasma concentrations of glucose and insulin, glucose fluxes, and insulin fluxes.}
\label{fig:S2008}
\end{figure}

Kovatchev et al. describe the glucose-insulin dynamics in detail \cite{UVA_1}. An overview is illustrated in Fig. \ref{fig:S2008}. The simulator with the mathematical model was accepted by the {\it Food and Drug Administration} (FDA) in 2008 as an in silico simulation platform. The model describes the relationship among plasma glucose and insulin concentration, glucose fluxes, and insulin fluxes. 
The glucose–insulin dynamics can be modeled by a 13-dimensional nonlinear ODE system. The state variables $x(t)\in\mathbb{R}^{13}$ are summarized in Table \ref{table:x}. The control input $u(t)\in\mathbb{R}$ is the insulin infusion rate (U/min) and the measurement $y_h\in\mathbb{R}$ denotes the glucose concentration measured by a CGM, which is called a {\it CGM value}. In addition, meal intake is modeled as a disturbance $d(t)\in\mathbb{R}$ that influences glucose mass in the stomach. For further details, the reader is referred to \cite{UVA_1,Meal_Simulation_Model}.
\begin{table}[t]
\centering
\caption{13-dimensional state variables of the glucose-insulin dynamics}
\label{table:x}
\begin{tabular}{l|l}
\hline
Variable & Description \\
\hline
$Q_{sto1}$ & Solid glucose mass in the stomach (mg) \\
$Q_{sto2}$ & Liquid glucose mass in the stomach (mg) \\
$Q_{gut}$ & Glucose mass in the intestine (mg) \\
$G_p$ & Glucose concentration in plasma (mg/kg) \\
$G_t$ & Glucose concentration in slowly equilibrating tissue (mg/kg) \\
$I_p$ &  Insulin concentration in plasma (pmol/kg) \\
$X$ &  Insulin action in the interstitial fluid (pmol/kg) \\
$I_1$ & Intermediate state of the delayed insulin signal (pmol/kg) \\
$I_d$ & Delayed insulin signal state (pmol/kg) \\
$I_l$ & Insulin concentration in the liver (pmol/kg) \\
$I_{sc1}$ & Subcutaneous insulin (first compartment) (pmol/kg) \\
$I_{sc2}$ & Subcutaneous insulin (second compartment) (pmol/kg) \\
$G_{sc}$ & Subcutaneous glucose concentration (mg/kg) \\
\hline
\end{tabular}
\end{table}
\subsection{Objective}
We design a networked digital controller that computes the insulin infusion rate based on CGM values, as shown in Fig. \ref{fig:ap_system}. The primary objective is to maintain the CGM value within the target range of 70--180 mg/dL while preventing hypoglycemia and severe hyperglycemia. The target range is clinically recommended to reduce the risks associated with both hypoglycemia and severe hyperglycemia \cite{TIR}. In addition, our goal is to reduce the number of communication events to improve energy efficiency. In this study, we consider a DRL-based event-triggered controller framework.

\section{DRL-based Event-Triggered Controller Design for Artificial Pancreas Systems}
To achieve the objective described in Section III-B, we employ two DRL algorithms: the {\it hierarchical event-triggered PPO} (H-ETPPO) algorithm \cite{ETC_DRL_1} as a baseline and the proposed {\it CGM-based event-triggered PPO} (CGM-ETPPO) algorithm. In both algorithms, the state of the environment $s_h\in\mathcal{S}$ is defined as a two-dimensional vector consisting of the CGM value and the previously applied insulin infusion rate. In RL, an {\it episode} is defined as a sequence of states, actions, and rewards generated through the interaction between an agent and an environment over a finite horizon. In this study, an episode ends when either the maximum number of steps $H$ is reached or the CGM value falls below a hypoglycemic threshold (10 mg/dL) or exceeds a hyperglycemic threshold (600 mg/dL). The termination of an episode is often represented by a binary variable $d_h$, which takes the value 1 when the episode ends and 0 otherwise.

\noindent {\it Remark}: The decision-making problem considered in this study is partially observable since the entire state of the system $x(t)$ is not directly accessible. Nevertheless, to focus on evaluating the proposed event-triggered mechanism in terms of control performance and communication efficiency, we adopt a simplified setting without incorporating observation histories or recurrent architectures. Despite this simplification, the proposed method achieves satisfactory performance, as shown in Section V.A. A more comprehensive treatment of partial observability is left for future work.

\subsection{Hierarchical Event-Triggered PPO}
In the H-ETPPO algorithm, the agent's action is defined as $a_h=[u_h\ e_h]^{\top}$, where $u_h\in\mathbb{R}$ denotes the control input (the insulin infusion rate) and $e_h\in\{0,1\}$ indicates whether the control input is updated. The policy is factorized into two components as follows: 
\begin{eqnarray}
    \pi_{\theta}(a_h|s_h)=\pi_{\theta}^{(e)}(e_h|s_h)\pi_{\theta}^{(u)}(u_h|s_h),\nonumber
\end{eqnarray}
where $\pi_{\theta}^{(e)}$ is a stochastic policy following a Bernoulli distribution and $\pi_{\theta}^{(u)}$ is a probability density function over continuous values. The probability ratio can be decomposed as follows:
\begin{eqnarray}
\rho_h(\theta)=\frac{\pi_{\theta}(a_h|s_h)}{\pi_{\theta_{\text{old}}}(a_h|s_h)}&=&\frac{\pi_{\theta}^{(e)}(e_h|s_h)}{\pi_{\theta_{\text{old}}}^{(e)}(e_h|s_h)}\cdot\frac{\pi_{\theta}^{(u)}(u_h|s_h)}{\pi_{\theta_{\text{old}}}^{(u)}(u_h|s_h)}\nonumber\\
&=:&\rho_{h}^{(e)}(\theta)\rho_{h}^{(u)}(\theta).\label{probability_rate_HETPPO}
\end{eqnarray}
To maintain the target range 70--180 mg/dL while reducing the communication frequency, we define the following reward function.
\begin{eqnarray}
R(s_h,e_h)=R^{(1)}(s_h)-\eta_e e_h, \label{H-ET_reward_func}
\end{eqnarray}
where
\begin{eqnarray}
    R^{(1)}(s_h)=\begin{cases}
        1, & 70\le y_h\le 180,\\
        0, & \text{otherwise},
    \end{cases}\label{reward_1}
\end{eqnarray}
and $\eta_e>0$ is a constant. The second term represents a penalty associated with updating the control input. 

The H-ETPPO algorithm is summarized in Algorithm \ref{alg:HETDRL}. Since the control input $u_h$ is not applied to the environment when $e_h = 0$, $\pi_{\theta}^{(u)}$ is updated using only the samples collected at the event-triggered time steps
\begin{eqnarray}
\mathcal{I} := \{ h \mid e_h = 1 \}.
\end{eqnarray}
The policy $\pi_{\theta}$ is updated to maximize the following objective using trajectories $(s_0,a_0,r_0,s_1,...,s_N)$ generated by the policy $\pi_{\theta_{\text{old}}}$.
\begin{eqnarray}
    J(\theta)=L_{\text{clip}}^{(e)}(\theta)+L_{\text{clip}}^{(u)}(\theta)+c_{\text{ent}}L_{\text{ent}}(\theta),\label{H_TEPPO_objective}
\end{eqnarray}
where
\begin{eqnarray}
&&L_{\text{clip}}^{(e)}(\theta)=\frac{1}{N}
\sum_{h=0}^{N-1}\min\Big(\rho_h^{(e)}(\theta)A^{\pi_{\theta_{\text{old}}}}(s_h,a_h),\nonumber \\
&&\hspace{1cm}\text{clip}(\rho_h^{(e)}(\theta),\, 1-\epsilon,\, 1+\epsilon)A^{\pi_{\theta_{\text{old}}}}(s_h,a_h)
\Big),\label{clip_objective_HETPPO_e}\\
&&L_{\text{clip}}^{(u)}(\theta)=
\frac{1}{|\mathcal{I}|}\sum_{h\in\mathcal{I}}\min\Big(\rho_h^{(u)}(\theta)A^{\pi_{\theta_{\text{old}}}}(s_h,a_h),\nonumber \\
&&\hspace{1cm}\text{clip}(\rho_h^{(u)}(\theta),\, 1-\epsilon,\, 1+\epsilon)A^{\pi_{\theta_{\text{old}}}}(s_h,a_h)
\Big),\label{clip_objective_HETPPO_u}
\end{eqnarray}
$L_{\text{ent}}(\theta)$ denotes the entropy bonus, and $c_{\text{ent}}>0$ is its coefficient. The estimation of the advantage value $A^{\pi_{\theta_{\text{old}}}}(s_h,a_h)$ and the update of the value function $V_{\phi}$ are the same as in the standard PPO.

\begin{algorithm}[tb]
\caption{Hierarchical Event-Triggered PPO}
\label{alg:HETDRL}
\begin{algorithmic}[1]
 \renewcommand{\algorithmicrequire}{\textbf{Input:}}
 \renewcommand{\algorithmicensure}{\textbf{Output:}}
 \STATE Initialize the policy $\pi_{\theta}$, the value function $V_{\phi}$, the buffer $\mathcal{D}\leftarrow\emptyset$, the set of event-triggered timesteps $\mathcal{I}\leftarrow\emptyset$. 
\FOR{episode$=1,...,E_{\max}$}
\STATE $h\leftarrow 0$, $s_0\sim\rho_0$, $d_0\leftarrow0$, $\tilde{u}\leftarrow0$.
\WHILE{$d_h\neq1$}
\STATE Determine whether to update insulin infusion rate $e_h\sim \pi_{\theta}^{(e)}(\cdot|s_h)$.
\IF{$e_h=1$}
\STATE Determine insulin infusion rate $u_h\sim \pi_{\theta}^{(u)}(\cdot|s_h)$.
\STATE Update $\tilde{u}\leftarrow u_h$.
\STATE Record the event-triggered time $\mathcal{I}\leftarrow\mathcal{I}\cup\{h\}$.
\ENDIF
\STATE Observe the next state $s_{h+1}$.
\STATE Receive the reward: $r_h=R(s_h,e_h)$.
\IF{$h+1=H$ or $s_{h+1}$ satisfies the terminal condition}
\STATE $d_{h+1}\leftarrow1$.
\ELSE
\STATE $d_{h+1}\leftarrow0$
\ENDIF
\STATE Store $(s_h,a_h,\tilde{u},r_h,s_{h+1},d_{h+1})$ in $\mathcal{D}$.
\IF{$|\mathcal{D}|= N$}
\STATE Compute $\hat{A}_h$ for all samples using GAE.
\STATE Update $\theta$ by Eq. (\ref{H_TEPPO_objective}).
\STATE Update $\phi$ by Eq. (\ref{PPO_critic}).
\STATE $\mathcal{D}\leftarrow \emptyset$, $\mathcal{I}\leftarrow\emptyset$.
\ENDIF
\STATE $h\leftarrow h+1$.
\ENDWHILE
\ENDFOR
 \end{algorithmic} 
 \end{algorithm}

\subsection{CGM-based Event-Triggered PPO}
Instead of learning the triggering mechanism as part of the policy $\pi_{\theta}^{(e)}$, we adopt a rule-based triggering mechanism based on the magnitude of changes in CGM values. This reduces the complexity of the learning problem by avoiding joint optimization of the control input and the triggering rule. In the proposed method, the agent makes decisions about insulin infusion rate only at time instants that satisfy a predefined condition, rather than at every time step $h = 0,1,\ldots,H$. Let $h_k$ denote the $k$-th decision-making time. We define the event-triggering interval, i.e., the duration for which the control input determined at the $k$-th update is held as follows:
\begin{eqnarray}
    \tau_{k}=\min\{i\in\mathbb{N}\ |\ |y_{h_{k}+i}-y_{h_k}|\ge\eta_{k}\},
\end{eqnarray}
where $\eta_k$ denotes the event-triggering threshold for the $(k+1)$-th update of the insulin infusion rate and is referred to as the $k$-th {\it CGM-threshold}. We propose two types of schemes for the threshold $\eta_k$ as follows:
\begin{itemize}
\item {\it Fixed threshold scheme}: The threshold is set to a predefined constant ($\eta_{k}\equiv\bar{\eta}$).  
\item {\it Variable threshold scheme}: The threshold is treated as a variable. In this case, the agent determines both the insulin infusion rate \(u_k\) and the threshold \(\eta_k\) as its action.
\end{itemize}
Although the variable threshold scheme increases the complexity of the learning problem, it allows the triggering condition to be adaptively adjusted.

In the proposed method, since actions persist for variable durations, the decision-making problem is naturally formulated as an SMDP, which generalizes the standard MDP \cite{SMDP}. In the SMDP formulation, training is performed using the following experience defined at the $k$-th decision-making: $(s_{k},a_{k},R_{k},s_{k+1})$, where $s_k=s_{h_k}$, $a_{k}=a_{h_k}$, $s_{k+1}=s_{h_k+\tau_{k}}$, and 
\begin{eqnarray}
R_{k}=\sum_{i=0}^{\tau_k-1}\gamma^{i}R(s_{h_k+i},a_{k}).\label{SMDP_rewards}
\end{eqnarray}
The reward function is defined as follows:
\begin{eqnarray}
R(s_h,a_h)=R^{(1)}(s_h)+R^{(2)}(s_h,a_h), \label{reward_func_SMDP}
\end{eqnarray}
where $R^{(1)}$ is Eq. (\ref{reward_1}), 
\begin{eqnarray}
    R^{(2)}(s_h, a_h)=\begin{cases}
        \frac{\ell_h-c}{C} & 70\le y_h\le 180,\\
        0 & \text{otherwise},
    \end{cases}\label{reward_2}
\end{eqnarray}
where $c$ and $C$ are constants that determine the offset and scaling of the reward, respectively. Let $\ell_h = h - h_{\text{pre}}$, where $h_{\text{pre}}$ denotes the most recent decision-making time at which the insulin infusion rate was updated.

We extend the PPO algorithm \cite{PPO} to the SMDP setting using extended GAE \cite{EGAE}. Our proposed algorithm is summarized in Algorithm \ref{alg:CGMETDRL}, focusing on the variable threshold scheme. The fixed threshold case is a special case obtained by fixing $\eta_k$ to a constant. From line 5 to line 20, an agent explores and collects experiences based on the current policy $\pi_{\theta}$. After that, from line 22 to line 25, the agent updates its policy using collected experiences $(s_0,a_0,R_0,s_1,\ldots,s_N)$. Similar to the standard PPO algorithm, the advantage function is estimated using the following recursive formulation.
\begin{eqnarray}
\hat{A}_{k}^{\text{SMDP}}&=&\delta_{k}^{\text{SMDP}}+\gamma^{\tau_k}\lambda(1-d_{k+1})\hat{A}_{k+1}^{\text{SMDP}},\nonumber\\
    \hat{A}_{N}^{\text{SMDP}}&=&0\nonumber,    
\end{eqnarray}
where $N$ denotes the size of $\mathcal{D}$ and 
\begin{eqnarray}
\delta_{k}^{\text{SMDP}}=R_{k}+\gamma^{\tau_k}(1-d_{k+1})V_{\phi}(s_{k+1})-V_{\phi}(s_{k}).\nonumber
\end{eqnarray}
$d_{k+1}\in\{0,1\}$ is a terminal flag, where $d_{k+1}=1$ if $h_{k+1}$ corresponds to the end of an episode, and $d_{k+1}=0$ otherwise. Therefore, $(1-d_{k+1})$ masks the bootstrap term $V_{\phi}(s_{k+1})$, ensuring that it is used only for non-terminal transitions. Using $\hat{A}_{k}^{\text{SMDP}}$ obtained above, the clipped surrogate objective can be defined in the same form as in the standard MDP. We use the following objective.
\begin{eqnarray}
    J^{\text{SMDP}}(\theta)=L_{\text{clip}}^{\text{SMDP}}(\theta)+c_{\text{ent}}L_{\text{ent}}^{\text{SMDP}}(\theta),\label{SMDP_objective}
\end{eqnarray}
where
\begin{eqnarray}
&&L_{\text{clip}}^{\text{SMDP}}(\theta)=
\frac{1}{N}\sum_{k=0}^{N-1}\min\Big(\rho_k(\theta)\hat{A}_{k}^{\text{SMDP}},\nonumber \\
&&\hspace{1cm}\text{clip}(\rho_k(\theta),\, 1-\epsilon,\, 1+\epsilon)\hat{A}_{k}^{\text{SMDP}}
\Big),\label{SMDP_clip_objective}
\end{eqnarray}
$L_{\text{ent}}^{\text{SMDP}}(\theta)$ denotes an entropy bonus, and $c_{\text{ent}}>0$ is its coefficient. The value function is trained by minimizing the following mean squared error.
\begin{eqnarray}
    J^{\text{SMDP}}(\phi)=\frac{1}{N}\sum_{k=0}^{N-1}\left(V_{\phi}(s_k)-\hat{G}_{k}\right)^2,\label{SMDP_critic_objective}
\end{eqnarray}
where $\hat{G}_{k}$ denotes the empirical return or its approximation.

In the proposed method, event triggering is determined by a clear and interpretable criterion, namely that the change in the glucose level exceeds a predefined threshold. This enhances interpretability compared to hierarchical approaches, where event triggering is determined by a black-box policy. Such interpretability is a key advantage for safety assessment and clinical acceptability in real-world deployment.
 
\begin{algorithm}[tb]
\caption{CGM-based Event-Triggered PPO}
\label{alg:CGMETDRL}
\begin{algorithmic}[1]
 \renewcommand{\algorithmicrequire}{\textbf{Input:}}
 \renewcommand{\algorithmicensure}{\textbf{Output:}}
 \STATE Initialize the policy $\pi_{\theta}$, the value function $V_{\phi}$ and the buffer $\mathcal{D}\leftarrow\emptyset$. 
\FOR{episode$=1,...,E_{\max}$}
\STATE $h\leftarrow0$, $k\leftarrow0$, $s_0\sim\rho_0$, $d_0\leftarrow0$. 
\WHILE{$d_{k}\neq1$ and $h<H$}
\STATE $s_k\leftarrow s_h$, $h_k\leftarrow h$, $y_{\text{start}}\leftarrow y_{h}$.
\STATE Determine the action $a_{k}=[u_k\ \eta_{k}]\sim\pi_{\theta}(\cdot|s_{k})$
\STATE $R_k\leftarrow0$, $\tau_k\leftarrow0$.
\REPEAT
    \STATE Observe the next state $s_{h+1}$.
    \STATE Receive the reward: $r_h=R(s_h,a_k)$.
    \STATE $h\leftarrow h+1$.
    \STATE $\tau_{k}\leftarrow \tau_{k}+1$, $R_{k}\leftarrow R_{k}+\gamma^{\tau_{k}-1}r_h$
\UNTIL{$h\ge H$ {\bf or} $s_h$ satisfies the terminal condition {\bf or} $|y_{h}-y_{\text{start}}|\ge\eta_k$}
\STATE $s_{k+1}\leftarrow s_{h}$.
\IF{$h\ge H$ or $s_{k+1}$ satisfies the terminal condition}
\STATE $d_{k+1}\leftarrow1$.
\ELSE
\STATE $d_{k+1}\leftarrow0$
\ENDIF
\STATE Store $(s_k,a_k,R_k,\tau_k,s_{k+1},d_{k+1})$ in $\mathcal{D}$.
\IF{$|\mathcal{D}|= N$}
\STATE Compute $\hat{A}_{k}^{\text{SMDP}}$ based on $V_{\phi}$ by extended GAE.
\STATE Update $\theta$ by Eq.\ (\ref{SMDP_objective}),
\STATE Update $\phi$ by Eq.\ (\ref{SMDP_critic_objective}).
\STATE $\mathcal{D}\leftarrow \emptyset$
\ENDIF
\STATE $k\leftarrow k+1$.
\ENDWHILE
\ENDFOR
 \end{algorithmic} 
 \end{algorithm}

\section{Numerical Experiments}
This section presents numerical experiments to evaluate the effectiveness of the proposed method. In these experiments, we use the Python library {\it Simglucose} \cite{Simglucose} based on the UVA/Padova T1D simulator  \cite{UVA_1,UVA_2}. Simglucose provides a simulation platform that includes virtual patients with diverse physiological characteristics. In this study, we conduct experiments on 10 virtual adult patients (adult\#001--adult\#010). The glucose-insulin dynamics are simulated using the Runge–Kutta method with a time step of 1 min. In this library, we can use three types of sensors ({\it Dexcom}, {\it GuardianRT}, and {\it Navigator}) and two types of pumps ({\it Cozmo} and {\it Insulet}), which differ in their measurement accuracy, sampling frequency, and actuation characteristics. In this study, we conduct experiments using the Dexcom sensor and the Insulet pump. The insulin infusion rate is constrained to the range of 0--0.15 U/min. Bolus administration is disabled, and only basal insulin is considered as the control input $u(t)$. In addition, meal-induced disturbances occur stochastically, with up to six meals per day. When a meal occurs, the solid glucose mass in the stomach $Q_{sto1}(t)$ increases at a rate of 5 g/min until the predefined meal amount is fully delivered to the stomach. Details of meal timing and amounts are provided in Appendix A.

In each experiment, the initial state vector $x(0)\in\mathbb{R}^{13}$ is set to the default values specified in the Simglucose parameter file, and during training, only $G_p$, $G_t$, and $G_{sc}$ are sampled from a Gaussian distribution $\mathcal{N}(\mu,(0.1\mu)^2)$, where $\mu$ denotes the default value of each parameter. The sampling period of the Dexcom sensor, denoted by $\Delta$, is 3 min. The CGM measurement provided by the sensor is 
\begin{eqnarray}
    y_{h}=\frac{G_{sc}(h\Delta)}{V_g}+\epsilon_h,\nonumber
\end{eqnarray}
where $V_g$ denotes the volume of glucose distribution (dL/kg), which is patient-dependent. $\epsilon_h$ denotes measurement noise sampled from a temporally correlated non-Gaussian process implemented in Simglucose. The length of each episode is set to 48 hours ($H=960$). An episode is terminated early if the blood glucose level deviates significantly from the target range, i.e., hypoglycemia ($y_h<10$) or severe hyperglycemia ($y_h>600$). For each patient, a policy is trained through trial and error over 2,000 episodes. All updates of the DNNs were performed using mini-batch learning with 10 epochs and a mini-batch size of 128. After training, the policy is evaluated on 5 fixed test scenarios using the following metrics.
\begin{itemize}
\item Episode completion fraction (ECF): 
\begin{eqnarray}
    \text{ECF}\text{[\%]}=100\cdot\frac{T}{H}, \label{ECF}
\end{eqnarray}
where $T$ denotes the number of time steps until the end of the episode. ECF represents the ratio of the actual duration of the episode to the maximum episode length $H$. In this study, training is deemed unsuccessful when the ECF falls below 90\%.
\item Time in Range (TIR): 
\begin{eqnarray}
    \text{TIR}[\%]=100\cdot\frac{1}{H}\sum_{h=1}^{T}\mathbb{I}(70\le y_{h}\le 180),\label{TIR}
\end{eqnarray}
where $\mathbb{I}$ denotes an indicator function, which takes the value 1 if the condition is satisfied and 0 otherwise. TIR represents the percentage of time during which CGM measurements remain within the target range. A TIR of at least 70\% is generally considered desirable \cite{TIR}.
\item Action Update Reduction Rate (AURR): 
\begin{eqnarray}
    \text{AURR}[\%]=100\cdot\left(1-\frac{(H-T)+K}{H}\right),\label{AURR}
\end{eqnarray}
where $K$ denotes the number of updates of the insulin infusion. AURR quantifies the reduction in insulin infusion updates relative to the maximum number of possible updates.  
\end{itemize}
Each metric is averaged over five fixed scenarios. The procedure is repeated four times with different random seeds to account for stochasticity in training, and the results are reported as the mean and standard deviation across runs.

\subsection{Comparison with Existing Methods}
We compare the following controller design methods.
\begin{itemize}
\item PID
\item H-ETPPO
\item CGM-ETPPO
\end{itemize}
The gains of PID controllers are determined via a grid search, as detailed in Appendix B. In H-ETPPO, we evaluate three settings of $\eta_e \in \{0.0, 0.1, 0.5\}$ in the reward function (\ref{H-ET_reward_func}). In CGM-ETPPO, we employ the variable-threshold scheme, where the threshold is constrained such that $\eta_k\in[15, 25]$. In the reward function (\ref{reward_2}), the constants are set to $c = 5$ and $C = 10$. The hyperparameters for H-ETPPO and CGM-ETPPO are summarized in Table \ref{table:hypara}. The architectures of the DNNs are summarized in Tables \ref{table:H-ETPPO} and \ref{table:CGM-ETPPO}. 

The ECF values corresponding to each design method are presented in Table \ref{table:ecf_results}. The results show that both PID and CGM-ETPPO achieve ECF values of 90\% or higher across all patients. In contrast, H-ETPPO results in unsuccessful training for most patients as indicated by ECF values below 90\%. This may be attributed to the increased complexity of the learning task, which involves jointly learning the insulin infusion rate and its update timing.

\begin{table}[t]
\centering
\caption{hyperparameters}
\label{table:hypara}
\begin{tabular}{l|l}
\hline
Parameter & Value \\
\hline
Discount Rate $\gamma$ & 0.99\\
GAE Parameter $\lambda$ & 0.95\\
Clipping Parameter $\epsilon$ & 0.2\\ 
Entropy Coefficient $c_{\text{ent}}$ & 0.01\\
Buffer Size $N$ & 512\\
Adam Learning Rate $\alpha$ & 3e-4\\
\hline
\end{tabular}
\end{table}

\begin{table*}[t]
\centering
\caption{Neural network architectures used in H-ETPPO}
\label{table:H-ETPPO}
\begin{tabular}{l|l|l|l|l}
\hline
Network & Architecture & Hidden Units & Activation & Output \\
\hline
Actor & MLP (2 layers) & 64, 64 & Tanh & Multi-head policy: Gaussian (insulin mean and log standard deviation), Bernoulli (event logit) \\
Critic & MLP (2 layers) & 64, 64 & Tanh & State value (linear output) \\
\hline
\end{tabular}
\end{table*}

\begin{table*}[t]
\centering
\caption{Neural network architectures used in PPO and CGM-ETPPO}
\label{table:CGM-ETPPO}
\begin{tabular}{l|l|l|l|l}
\hline
Network & Architecture & Hidden Units & Activation & Output \\
\hline
Actor & MLP (2 layers) & 64, 64 & Tanh & Gaussian policy (mean and log standard deviation) \\
Critic & MLP (2 layers) & 64, 64 & Tanh & State value (linear output) \\
\hline
\end{tabular}
\end{table*}

\begin{table*}[t]
\centering
\caption{Episode Completion Fraction (ECF, \%) across patients, comparing PID, H-ETPPO, and CGM-ETPPO.}
\label{table:ecf_results}
\begin{tabular}{l|ccccc}
\hline
Patient & PID & H-ETPPO ($\eta_{e}=0.0$) & H-ETPPO ($\eta_{e}=0.1$) & H-ETPPO ($\eta_{e}=0.5$) & CGM-ETPPO \\
\hline
adult\#001  & $\bf{100.00 \pm 0.00}$ & $24.83 \pm 4.67$ & $95.21 \pm 8.29$ & $68.87 \pm 33.20$ & $\bf{100.00 \pm 0.00}$   \\
adult\#002 & $\bf{100.00 \pm 0.00}$ & $38.65 \pm 19.50$ & $28.20 \pm 3.06$ & $44.48 \pm 32.11$ & $\bf{100.00 \pm 0.00}$   \\
adult\#003 & $\bf{100.00 \pm 0.00}$ & $48.38 \pm 37.76$ & $9.05 \pm 2.45$ & $33.02 \pm 38.85$ & $\bf{100.00 \pm 0.00}$  \\
adult\#004 & $\bf{100.00 \pm 0.00}$ & $78.75 \pm 36.81$ & $64.16 \pm 38.31$ & $45.96 \pm 24.67$ & $\bf{100.00 \pm 0.00}$   \\
adult\#005 & $\bf{100.00 \pm 0.00}$ & $40.43 \pm 34.64$ & $68.87 \pm 54.02$ & $40.67 \pm 34.90$ & $\bf{100.00 \pm 0.00}$   \\
adult\#006 & $98.88 \pm 1.95$ & $33.46 \pm 35.95$ & $32.81 \pm 38.79$ & $18.33 \pm 5.87$ & $\bf{100.00 \pm 0.00}$   \\
adult\#007 & $\bf{100.00 \pm 0.00}$ & $60.40 \pm 39.92$ & $35.84 \pm 18.09$ & $44.15 \pm 38.21$ & $\bf{100.00 \pm 0.00}$  \\
adult\#008 & $\bf{100.00 \pm 0.00}$ & $10.14 \pm 0.11$ & $10.29 \pm 0.47$ & $56.50 \pm 43.55$ & $\bf{100.00 \pm 0.00}$  \\
adult\#009 & $94.99 \pm 6.75$ & $25.83 \pm 7.19$ & $52.99 \pm 26.98$ & $40.77 \pm 31.13$ & $\bf{100.00 \pm 0.00}$  \\
adult\#010 & $\bf{100.00 \pm 0.00}$ & $40.29 \pm 34.48$ & $40.25 \pm 34.50$ & $33.36 \pm 19.53$ & $\bf{100.00 \pm 0.00}$  \\
\hline
\end{tabular}
\end{table*}

The TIR values for PID and CGM-ETPPO are shown in Table \ref{table:tir_results}. The results show that both methods achieve comparable TIR values across all patients. In contrast, CGM-ETPPO achieves AURR values exceeding 90\% while maintaining control performance comparable to that of a well-tuned PID controller, as shown in Table \ref{table:aurr_results}. To improve TIR compared to PID, it is desirable to learn a policy that uses the history of CGM values, for example by using a recurrent neural network, to predict deviations in CGM values from the target range. Addressing partial observability in CGM-ETPPO through such approaches is left for future work.

\begin{figure}[htbp]
\centering
\includegraphics[width=6.5cm]{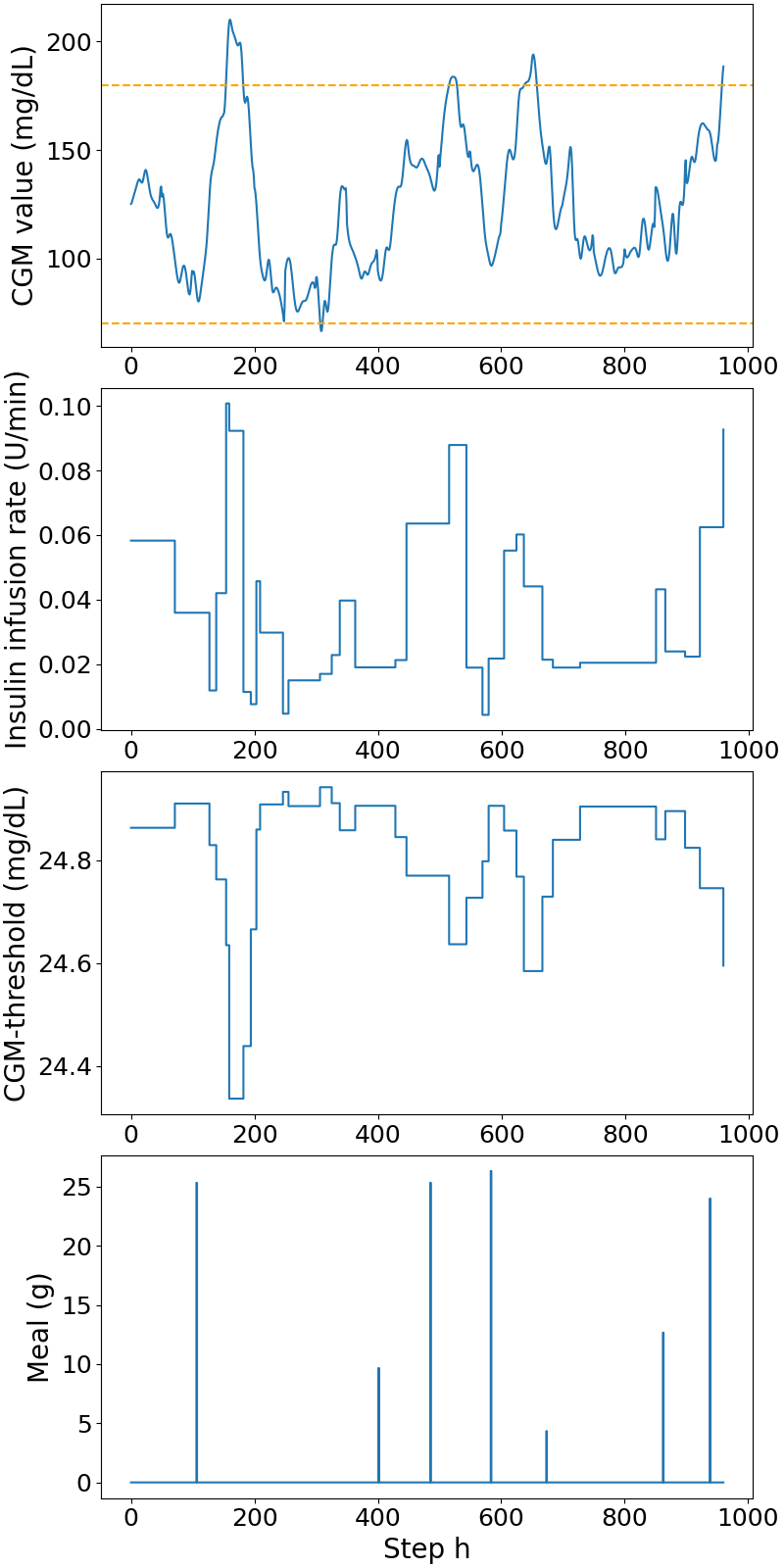}
\caption{Time responses under the policy learned by CGM-ETPPO for adult\#002. The first, second, and third plots show CGM value, insulin infusion rate, and variable CGM-threshold, respectively. The fourth plot indicates meal timing.}
\label{fig:tr_cgm_etppo}
\end{figure}

An example of insulin infusion under the policy learned by CGM-ETPPO is shown in Fig. \ref{fig:tr_cgm_etppo}. It is observed that, when the CGM measurements fall outside the target range due to meal intake, the policy increases the insulin infusion rate to prevent further deviation and bring the glucose level back into the target range. 

\begin{table}[t]
\centering
\caption{Time in Range (TIR, \%) across patients, comparing PID and CGM-ETPPO.}
\label{table:tir_results}
\begin{tabular}{l|cc}
\hline
Patient & PID & CGM-ETPPO \\
\hline
adult\#001  & $65.11 \pm 1.17$ & $66.89 \pm 4.39$ \\
adult\#002 & $83.88 \pm 1.14$ & $84.56 \pm 1.33$   \\
adult\#003 & $77.23 \pm 1.30$ & $77.84 \pm 1.47$  \\
adult\#004 & $68.06 \pm 0.67$ & $66.30 \pm 3.77$ \\
adult\#005 & $75.27 \pm 1.08$ & $79.99 \pm 0.45$ \\
adult\#006 & $56.01 \pm 1.25$ & $52.89 \pm 1.25$  \\
adult\#007 & $57.73 \pm 1.91$ & $74.35 \pm 1.30$ \\
adult\#008 & $91.61 \pm 1.07$ & $90.88 \pm 2.76$ \\
adult\#009 & $51.36 \pm 3.62$ & $52.41 \pm 5.09$ \\
adult\#010 & $62.17 \pm 1.52$ & $63.49 \pm 2.94$  \\
\hline
\end{tabular}
\end{table}

\begin{table}[t]
\centering
\caption{Action Update Reduction Rate (AURR, \%) of CGM-ETPPO across patients}
\label{table:aurr_results}
\begin{tabular}{l|c}
\hline
Patient & CGM-ETPPO \\
\hline
adult\#001  & $96.45 \pm 0.36$  \\
adult\#002 & $96.37 \pm 0.25$    \\
adult\#003 & $95.82 \pm 0.16$   \\
adult\#004 & $93.53 \pm 0.27$  \\
adult\#005 & $96.18 \pm 0.08$  \\
adult\#006 & $94.13 \pm 0.06$   \\
adult\#007 & $95.46 \pm 0.26$  \\
adult\#008 & $97.02 \pm 0.21$  \\
adult\#009 & $95.36 \pm 0.21$  \\
adult\#010 & $95.11 \pm 0.20$   \\
\hline
\end{tabular}
\end{table}

\subsection{Analysis of the Variable Threshold with Respect to CGM Values}
\begin{figure}[htbp]
\centering
\includegraphics[width=8.0cm]{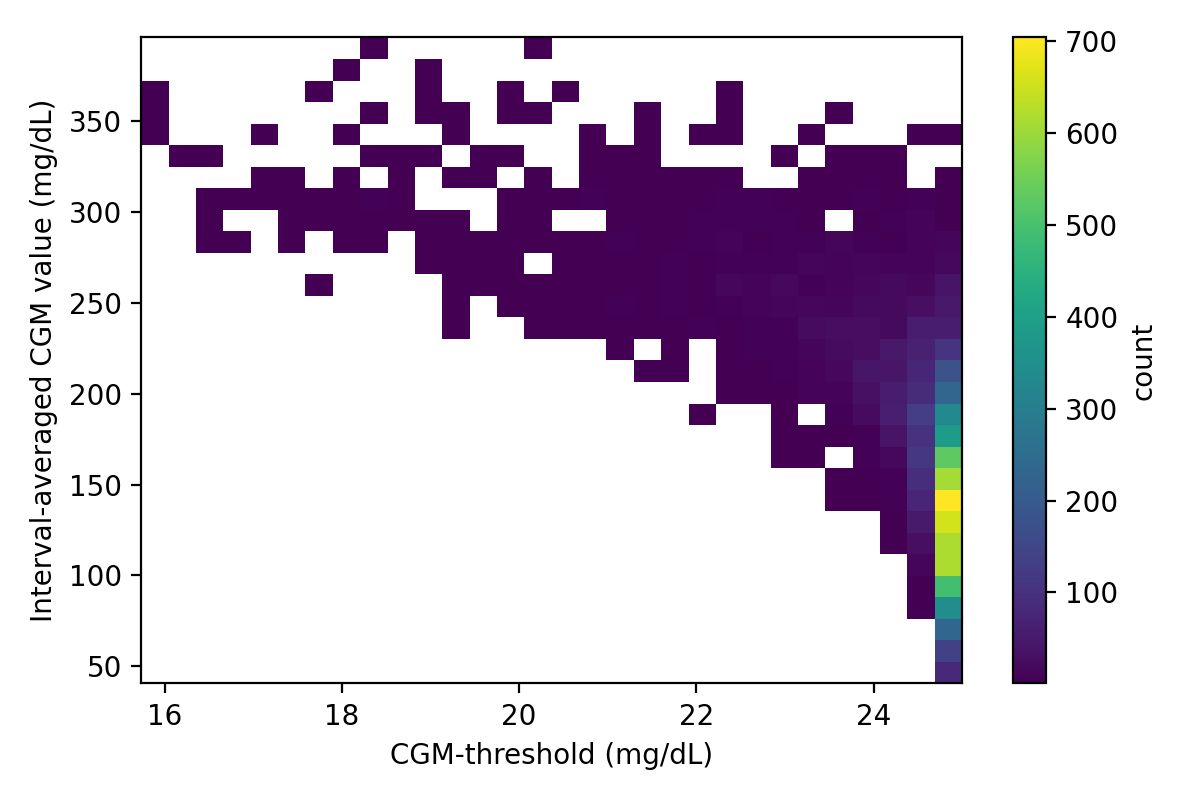}
\caption{Histograms of the interval-averaged CGM values and the corresponding variable thresholds obtained from the simulation results. White regions indicate the absence of data.}
\label{fig:hist}
\end{figure}

We analyze the behavior of the variable CGM-threshold with respect to CGM values. To examine this relationship, we define the {\it interval-averaged CGM value} as follows:
\begin{eqnarray}
    \bar{c}_{k}=\frac{1}{h_{k+1}-h_{k}}\sum_{h=h_k}^{h_{k+1}-1}y_{h},\nonumber
\end{eqnarray}
which represents the average CGM value over the interval between two insulin infusion rate updates. The histograms of the interval-averaged CGM values $\bar{c}_{k}$ and the corresponding CGM-threshold values $\eta_k$ are shown in Fig.\ \ref{fig:hist}. Under the reward term defined in Eq. (\ref{reward_2}), the reward increases as the update interval $h_{k+1}-h_{k}$ becomes longer. As a result, the learned policy tends to determine larger CGM-thresholds when the interval-averaged CGM value lies within the target range 70--180 mg/dL to suppress the frequency of insulin infusion updates. Indeed, as shown in Fig. \ref{fig:hist}, CGM-thresholds close to the upper bound (25 mg/dL) are more frequently selected. In contrast, under hyperglycemic conditions $y_h\ge 180$, a large threshold hinders timely updates of the insulin infusion rate, which may result in delayed glucose regulation. Consequently, smaller CGM-thresholds tend to be selected as the CGM value increases. As shown in Fig. \ref{fig:tr_cgm_etppo}, the CGM-threshold decreases as the CGM value increases. Meanwhile, even under hyperglycemic conditions, when the glucose level is below 200 mg/dL, the CGM-threshold is concentrated around 22--25 mg/dL. This suggests that a nearly constant CGM-threshold may still yield effective control performance. In the next section, we further evaluate a fixed threshold scheme.

\subsection{Comparison between fixed and variable threshold schemes}
We evaluate the performance of fixed and variable threshold schemes in which only the insulin infusion rate is treated as the agent's action. The fixed CGM-threshold is selected from the set $\bar{\Theta}=\{15,20,25\}$. The performance metrics, including ECF, TIR, and AURR, are summarized in Tables \ref{table:ecf_results_secB}, \ref{table:tir_results_secB}, and \ref{table:aurr_results_secB}, respectively. For all patients except adult\#009, the ECF values were consistently 100\% and no notable differences were observed in the other performance metrics. These results indicate that, even with a fixed threshold, the learned policy can suppress fluctuations and achieve improved performance across the evaluation metrics by maximizing the reward $R^{(2)}$. This may be attributed to the fact that a fixed threshold is sufficient for most patients to allow effective insulin infusion.

\begin{figure}[htbp]
\centering
\includegraphics[width=6.5cm]{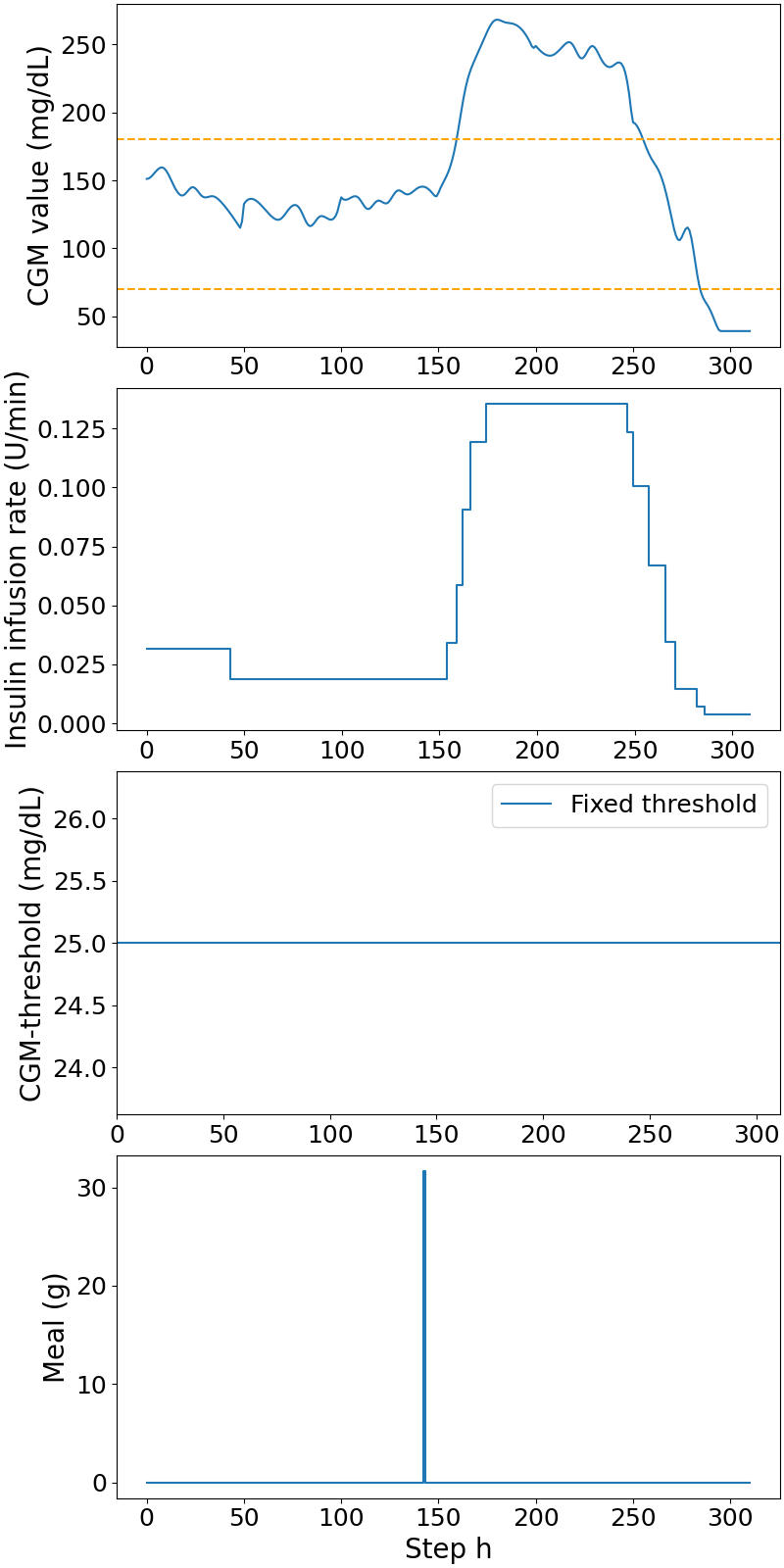}
\caption{Time responses under the policy learned by CGM-ETPPO with the fixed threshold scheme ($\eta\equiv25$) for adult\#009. The first, second, and third plots show CGM value, insulin infusion rate, and the CGM-threshold, respectively. The fourth plot indicates the meal timing. The episode terminates early due to hypoglycemia.}
\label{fig:tr_cgm_etppo_fixed}
\end{figure}

\begin{figure}[htbp]
\centering
\includegraphics[width=6.5cm]{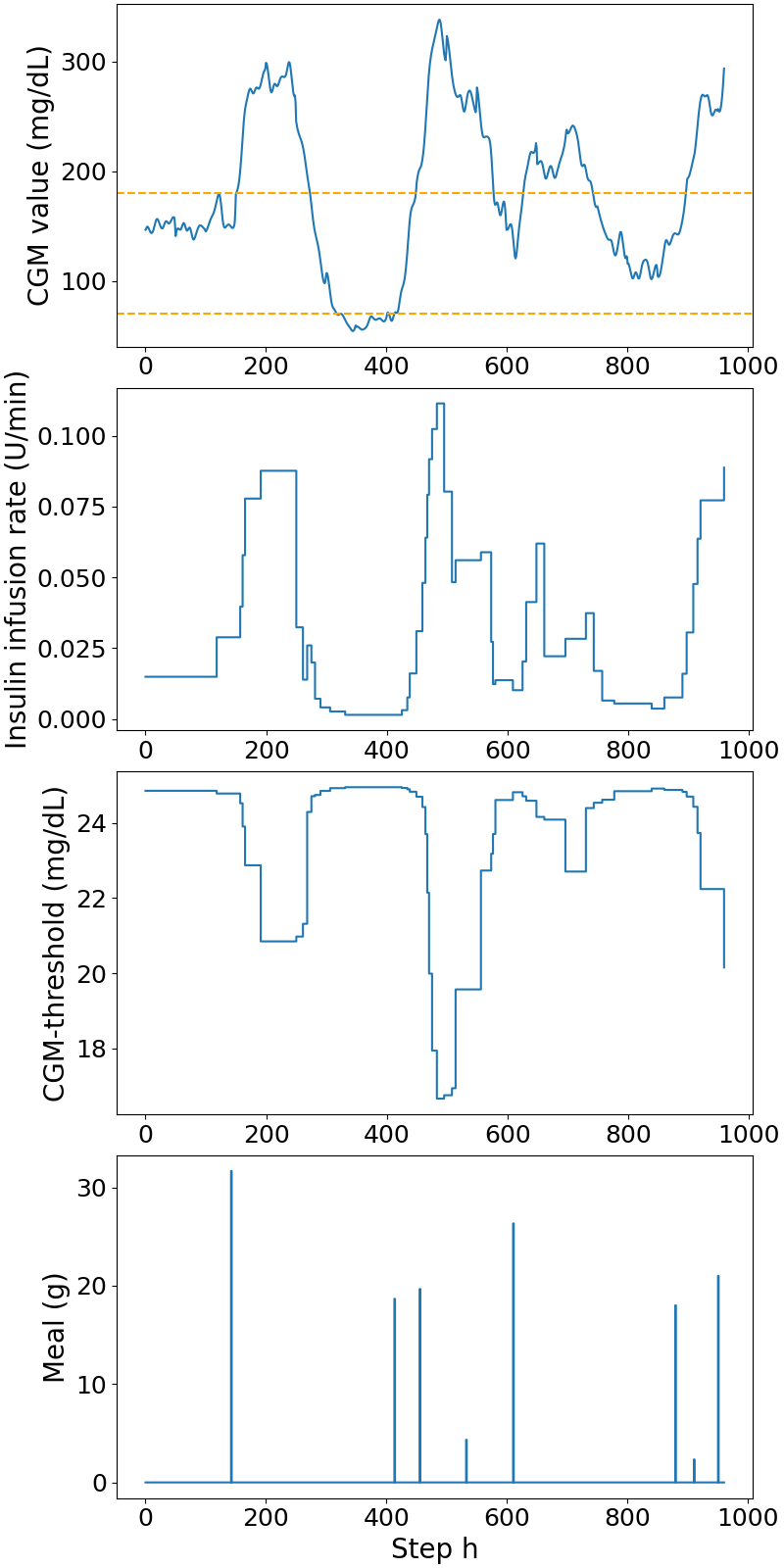}
\caption{Time responses under the policy learned by CGM-ETPPO with the variable threshold scheme for adult\#009. The first, second, and third plots show CGM value, insulin infusion rate, and the variable CGM-threshold, respectively. The fourth plot indicates the meal timing.}
\label{fig:tr_cgm_etppo_variable}
\end{figure}

Nevertheless, there are cases where a variable threshold scheme is particularly effective. For example, for adult\#009, while the fixed threshold scheme does not achieve 100\% ECF, the variable threshold scheme shows improved performance. Examples of insulin infusion for adult\#009 with a fixed threshold scheme ($\eta\equiv25$) and with a variable threshold scheme are shown in Fig.~\ref{fig:tr_cgm_etppo_fixed} and Fig.~\ref{fig:tr_cgm_etppo_variable}, respectively. When the CGM-threshold is fixed, insulin infusion may be insufficient to adequately respond to glucose fluctuations, resulting in hypoglycemic events and early termination of the episode.

While fixing the threshold can simplify the learning problem, it may limit the ability to achieve effective control depending on the individual patient dynamics.

\begin{table*}[t]
\centering
\caption{Episode Completion Fraction (ECF, \%) across patients, comparing the fixed threshold scheme and the variable threshold scheme}
\label{table:ecf_results_secB}
\begin{tabular}{l|ccc|c}
\hline
Patient & $\eta=15$ & $\eta=20$ & $\eta=25$ & CGM-ETPPO ($\eta\in[15,25]$)\\
\hline
adult\#001  & $100.00 \pm 0.00$ & $100.00 \pm 0.00$ & $100.00 \pm 0.00$ &  $100.00 \pm 0.00$ \\
adult\#002 & $100.00 \pm 0.00$ & $100.00 \pm 0.00$ & $100.00 \pm 0.00$ & $100.00 \pm 0.00$  \\
adult\#003 & $100.00 \pm 0.00$ & $100.00 \pm 0.00$ & $100.00 \pm 0.00$ & $100.00 \pm 0.00$ \\
adult\#004 & $100.00 \pm 0.00$ & $100.00 \pm 0.00$ & $100.00 \pm 0.00$ & $100.00 \pm 0.00$ \\
adult\#005 & $100.00 \pm 0.00$ & $100.00 \pm 0.00$ & $100.00 \pm 0.00$ & $100.00 \pm 0.00$ \\
adult\#006 & $100.00 \pm 0.00$ & $100.00 \pm 0.00$ &$100.00 \pm 0.00$& $100.00 \pm 0.00$ \\
adult\#007 & $100.00 \pm 0.00$ & $100.00 \pm 0.00$ & $100.00 \pm 0.00$ & $100.00 \pm 0.00$ \\
adult\#008 & $100.00 \pm 0.00$ & $100.00 \pm 0.00$ & $100.00 \pm 0.00$ & $100.00 \pm 0.00$ \\
adult\#010 & $100.00 \pm 0.00$ & $100.00 \pm 0.00$ & $100.00 \pm 0.00$ & $100.00 \pm 0.00$ \\
\hline
adult\#009 & $55.29 \pm 19.93$ & $71.79 \pm 29.41$ & $93.39 \pm 11.46$ & $\mathbf{100.00 \pm 0.00}$ \\
\hline
\end{tabular}
\end{table*}

\begin{table*}[t]
\centering
\caption{Time in Range (TIR, \%) across patients, comparing the fixed threshold scheme and the variable threshold scheme}
\label{table:tir_results_secB}
\begin{tabular}{l|ccc|c}
\hline
Patient & $\eta=15$ & $\eta=20$ & $\eta=25$ & CGM-ETPPO ($\eta\in[15,25]$)\\
\hline
adult\#001  & $72.52 \pm 3.53$ & $68.96 \pm 5.58$ & $73.50 \pm 3.23$ &  $66.89 \pm 4.39$ \\
adult\#002 & $84.62 \pm 2.14$ & $84.96 \pm 2.53$ & $88.73 \pm 1.05$ & $84.56 \pm 1.33$  \\
adult\#003 & $78.41 \pm 1.52$ & $78.92 \pm 1.20$ & $76.68 \pm 1.08$ & $77.84 \pm 1.47$ \\
adult\#004 & $71.74 \pm 1.44$ & $70.22 \pm 0.76$ & $69.86 \pm 1.39$ & $66.30 \pm 3.77$ \\
adult\#005 & $78.77 \pm 2.78$ & $77.83 \pm 4.29$ & $77.72 \pm 1.73$ & $79.99 \pm 0.45$ \\
adult\#006 & $51.39 \pm 2.83$ & $53.04 \pm 0.88$ &$51.48 \pm 0.99$& $52.89 \pm 1.25$ \\
adult\#007 & $76.05 \pm 0.93$ & $75.04 \pm 2.02$ & $73.89 \pm 0.55$ & $74.35 \pm 1.30$ \\
adult\#008 & $92.64 \pm 0.68$ & $91.36 \pm 2.02$ & $90.26 \pm 1.73$ & $90.88 \pm 2.76$ \\
adult\#010 & $64.70 \pm 4.64$ & $61.61 \pm 1.86$ & $64.32 \pm 3.59$ & $63.49 \pm 2.94$ \\
\hline
adult\#009 & $33.96 \pm 8.33$ & $38.67 \pm 13.34$ & $47.92 \pm 5.71$ & $\mathbf{52.41 \pm 5.09}$ \\
\hline
\end{tabular}
\end{table*}

\begin{table*}[t]
\centering
\caption{Action Update Reduction Rate (AURR, \%) across patients, comparing the fixed threshold scheme and the variable threshold scheme}
\label{table:aurr_results_secB}
\begin{tabular}{l|ccc|c}
\hline
Patient & $\eta=15$ & $\eta=20$ & $\eta=25$ & CGM-ETPPO ($\eta\in[15,25]$)\\
\hline
adult\#001  & $93.38 \pm 0.12$ & $95.08 \pm 0.15$ & $96.55 \pm 0.20$ &  $96.45 \pm 0.36$ \\
adult\#002 & $92.97 \pm 0.31$ & $95.08 \pm 0.25$ & $96.31 \pm 0.14$ & $96.37 \pm 0.25$  \\
adult\#003 & $92.38 \pm 0.24$ & $94.35 \pm 0.20$ & $95.72 \pm 0.16$ & $95.82 \pm 0.16$ \\
adult\#004 & $90.05 \pm 0.39$ & $92.48 \pm 0.62$ & $94.17 \pm 0.17$ & $93.53 \pm 0.27$ \\
adult\#005 & $92.59 \pm 0.31$ & $94.90 \pm 0.43$ & $96.19 \pm 0.06$ & $96.18 \pm 0.08$ \\
adult\#006 & $89.92 \pm 0.48$ & $92.73 \pm 0.20$ &$94.18 \pm 0.18$& $94.13 \pm 0.06$ \\
adult\#007 & $92.01 \pm 0.32$ & $94.15 \pm 0.24$ & $95.47 \pm 0.14$ & $95.46 \pm 0.26$ \\
adult\#008 & $94.13 \pm 0.31$ & $95.91 \pm 0.33$ & $97.00 \pm 0.11$ & $97.02 \pm 0.21$ \\
adult\#010 & $91.30 \pm 0.40$ & $93.72 \pm 0.12$ & $95.33 \pm 0.19$ & $95.11 \pm 0.20$ \\
\hline
adult\#009 & $50.79 \pm 18.32$ & $67.63 \pm 27.69$ & $89.30 \pm 10.92$ & $\mathbf{95.36 \pm 0.21}$ \\
\hline
\end{tabular}
\end{table*}

\subsection{Comparison with Standard PPO}

We investigate the effectiveness of incorporating ETC. Specifically, we compare CGM-ETPPO with a fixed threshold of $\eta\equiv25$ and standard PPO, which is equivalent to CGM-ETPPO with the threshold fixed at $\eta\equiv0$. In this experiment, the reward function is defined by Eq. (\ref{reward_1}), which focuses solely on maintaining the CGM value within a target range. The DNN architectures and hyperparameters are the same as those used for CGM-ETPPO. 

The evaluation results for ECF are shown in Table \ref{table:ecf_results_secD}. It is observed that, while standard PPO fails to control for several patients, CGM-ETPPO consistently achieves ECF values above 90\% across all patients, except for adult\#009, where both methods exhibit limited performance. 
This improvement can be attributed to the event-triggered mechanism, which focuses control updates on significant changes in CGM values and reduces the effective degrees of freedom of the action space by limiting unnecessary updates, thereby simplifying the learning problem. 

\begin{table}[t]
\centering
\caption{Comparison of ECF between standard PPO and CGM-ETPPO across all patients.}
\label{table:ecf_results_secD}
\begin{tabular}{l|cc}
\hline
Patient & PPO & CGM-ETPPO ($\eta=25$) \\
\hline
adult\#001  & $49.03 \pm 32.18$ & $\bf{100.00 \pm 0.00}$   \\
adult\#002 & $73.48 \pm 29.48$ & $\bf{100.00 \pm 0.00}$   \\
adult\#003 & $81.79 \pm 14.00$ & $\bf{100.00 \pm 0.00}$  \\
adult\#004 & $77.33 \pm 39.26$ & $\bf{100.00 \pm 0.00}$   \\
adult\#005 & $30.18 \pm 18.41$ & $\bf{100.00 \pm 0.00}$   \\
adult\#006 & $77.64 \pm 38.74$ & $92.99 \pm 7.01$   \\
adult\#007 & $\bf{100.00 \pm 0.00}$ & $\bf{100.00 \pm 0.00}$  \\
adult\#008 & $36.23 \pm 37.31$ & $\bf{100.00 \pm 0.00}$  \\
adult\#009 & $53.67 \pm 29.76$ & $49.55 \pm 5.97$  \\
adult\#010 & $60.87 \pm 39.14$ & $\bf{100.00 \pm 0.00}$  \\
\hline
\end{tabular}
\end{table}

\section{Conclusion}
We proposed a DRL-based event-triggered control method for networked AP systems to maintain CGM values within the target range while reducing the communication frequency. We considered two PPO-based algorithms: H-ETPPO as a representative existing method and the proposed CGM-ETPPO. While H-ETPPO learns a policy that jointly determines the insulin infusion rate and the timing of updates, CGM-ETPPO learns a policy that determines only the insulin infusion rate, effectively decoupling the event-triggering mechanism from policy learning and thereby simplifying the learning problem. The resulting decision-making problem is naturally formulated as an SMDP, for which we extend a standard DRL algorithm to handle irregular decision intervals. Numerical experiments demonstrated the effectiveness of the proposed approach. We further analyzed the impact of the CGM threshold setting and the effect of event-triggered control on learning performance.

Several important directions remain for future work, as follows: First, partial observability should be addressed as the proposed method relies solely on CGM measurements. Incorporating mechanisms to handle partial observability, such as recurrent architectures, is an important direction. Second, the development of personalized control strategies based on meta-RL is promising. Such approaches could enable rapid adaptation to inter-patient variability with limited data. Finally, integrating PID control structures into the proposed framework, such as \cite{CIRL}, is of interest. Although the PID control structure imposes constraints on the degrees of freedom of the control input, it can reduce the complexity of learning and enable the learning of policies that are robust to inter-patient variability.

\appendices

\section{Meal Scenario Generation}

At the beginning of each episode, a stochastic meal scenario is generated. 
The scenario consists of a set of meal events characterized by their occurrence times and carbohydrate amounts.

Let $i \in \{1, \dots, 6\}$ denote the meal candidates (breakfast, snack 1, lunch, snack 2, dinner, snack 3). 
Each meal is included independently with probability $p_i$. If meal $i$ is selected, its occurrence time $t_i$ is sampled from a truncated normal distribution:
\begin{equation}
t_i \sim \mathrm{TruncNormal}(t_{\mu,i}, t_{\sigma,i}^2; t_{\mathrm{lb},i}, t_{\mathrm{ub},i}),\nonumber
\end{equation}
where $t_{\mu,i}$ and $t_{\sigma,i}$ denote the mean and standard deviation, and 
$t_{\mathrm{lb},i}$ and $t_{\mathrm{ub},i}$ are the lower and upper bounds, respectively. 
This truncation ensures that $t_i \in [t_{\mathrm{lb},i}, t_{\mathrm{ub},i}]$. 
The sampled time is rounded to the nearest minute, and $t_i$ is expressed in minutes.  The carbohydrate amount $m_i$ is sampled independently as
\begin{equation}
m_i \sim \mathcal{N}(m_{\mu,i}, m_{\sigma,i}^2),\nonumber
\end{equation}
and is clipped to ensure non-negativity:
\begin{equation}
m_i \leftarrow \max(m_i, 0).\nonumber
\end{equation}
The amount of carbohydrate $m_i$ is expressed in grams.

Through this procedure, each episode yields a stochastic daily meal pattern with variability in both meal timing and carbohydrate intake.

\begin{table}[t]
\centering
\caption{Parameters for meal time and carbohydrate distributions}
\begin{tabular}{l|cccccc}
\hline
Meal & $p_i$ & $t_{\mathrm{lb},i}$ & $t_{\mathrm{ub},i}$ & $t_{\mu,i}$ & $t_{\sigma,i}$ & $m_{\mu,i}$ ($m_{\sigma,i}$) \\
\hline
Breakfast & 0.95 & 300 & 540 & 420 & 60 & 45 (10) \\
Snack 1   & 0.30 & 540 & 600 & 570 & 30 & 10 (5) \\
Lunch     & 0.95 & 600 & 840 & 720 & 60 & 70 (10) \\
Snack 2   & 0.30 & 840 & 960 & 900 & 30 & 10 (5) \\
Dinner    & 0.95 & 960 & 1200 & 1080 & 60 & 80 (10) \\
Snack 3   & 0.30 & 1200 & 1380 & 1290 & 30 & 10 (5) \\
\hline
\end{tabular}
\label{tab:meal_params}
\end{table}

\section{Implementation of PID Control}
For the PID control baseline, we used the implementation provided in simglucose. The target glucose level was set to 112.5 mg/dL. The PID gains were selected via a grid search around the following default parameters: $K_p=0.001,\ K_i=0.00001,\ K_d=0.001$. As a result, $K_i=0.0$ and $K_d=0.01$ for all patients. The selected values $K_p$ are summarized in Table \ref{table:pid}. 

\begin{table}[t]
\centering
\caption{Selected proportional gains $K_p$ for each patient}
\begin{tabular}{l|c}
\hline
Patient & $K_p$ \\
\hline
adult\#001 & 0.0013 \\
adult\#002 & 0.0013 \\
adult\#003 & 0.0009 \\
adult\#004 & 0.0005 \\
adult\#005 & 0.0013 \\
adult\#006 & 0.0009 \\
adult\#007 & 0.0005 \\
adult\#008 & 0.0013 \\
adult\#009 & 0.0009 \\
adult\#010 & 0.0013 \\
\hline
\end{tabular}
\label{table:pid}
\end{table}









\begin{thebibliography}{99}
\bibitem{DM1}
    American Diabetes Association, ``Diagnosis and Classification of Diabetes Mellitus,'' {\it Diabetes Care}, vol. {\it 37}, no. {\it 1}, pp. S81--S90, 2014.

\bibitem{DM2}
    S. A. Antar et al., ``Diabetes Mellitus: Classification, Mediators, and Complications; A Gate to Identify Potential Targets for the Development of New Effective Treatments,'' {\it Biomedicine \& Pharmacotherapy}, vol. {\it 168}, 115734, 2023.

\bibitem{Care_DM}
American Diabetes Association Professional Practice Committee, ``2. Diagnosis and Classification of Diabetes: Standards of Care in Diabetes—2025,'' {\it Diabetes Care}, vol. {\it 48}, no. {\it 1}, pp. S27--S49, 2025.

\bibitem{CGM}
    The Juvenile Diabetes Research Foundation Continuous Glucose Monitoring Study Group, ``Continuous Glucose Monitoring and Intensive Treatment of Type 1 Diabetes,'' {\it New England Journal of Medicine}, vol. {\it 359}, no. {\it 14}, pp. 1464--1476, 2008.

\bibitem{AP1}
    C. K. Boughton et al., ``Hybrid Closed-Loop Glucose Control Compared with Sensor Augmented Pump Therapy in Older Adults with Type 1 Diabetes: An Open-Label Multicentre, Multinational, Randomised, Crossover Study,'' {\it The Lancet Healthy Longevity}, vol. {\it 3}, no. {\it 3}, pp. e135--e142, 2022.

\bibitem{AP2}
    J. Ware et al., ``Cambridge Hybrid Closed-Loop Algorithm in Children and Adolescents with Type 1 Diabetes: A Multicentre 6-Month Randomised Controlled Trial,'' {\it The Lancet Digital Health}, vol. {\it 4}, no. {\it 4}, pp. e245--e255, 2022.

\bibitem{PID_AP}
    G. M. Steil et al., ``Feasibility of Automating Insulin Delivery for the Treatment of Type 1 Diabetes,'' {\it Diabetes}, vol. {\it 55},\break   no. {\it 12}, pp. 3344--3350, 2006. 

\bibitem{UVA_1}
    B. P. Kovatchev et al., ``In Silico Preclinical Trials: A Proof of Concept in Closed-Loop Control of Type 1 Diabetes,'' {\it Journal of Diabetes Science and Technology}, vol. {\it 3}, no. {\it 1}, pp. 44--55, 2009.

\bibitem{UVA_2}
    C. D. Man et al., ``The UVA/PADOVA Type 1 Diabetes Simulator: New Features,'' {\it Journal of Diabetes Science and Technology}, vol. {\it 8}, no. {\it 1}, pp. 26--34, 2014.

\bibitem{MPC_AP1}
    L. Magni et al., ``Model Predictive Control of Type 1 Diabetes: An In Silico Trial,'' {\it Journal of Diabetes Science and Technology}, vol. {\it 1}, no. {\it 6}, pp. 804--812, 2007.

\bibitem{MPC_AP2}
    C. S. Hughes et al., ``Hypoglycemia Prevention via Pump Attenuation and Red-Yellow-Green “Traffic” Lights Using Continuous Glucose Monitoring and Insulin Pump Data,'' {\it Journal of Diabetes Science and Technology}, vol. {\it 4}, no. {\it 5}, pp. 1146--1155, 2010.

\bibitem{MPC_AP3}
    P. Soru et al., ``MPC Based Artificial Pancreas: Strategies for Individualization and Meal Compensation,'' {\it Annual Reviews in Control}, vol. {\it 36}, no. {\it 1}, pp. 118--128, 2012.

\bibitem{MPC_AP4}
    C. Toffanin et al., ``Artificial Pancreas: Model Predictive Control Design from Clinical Experience,'' {\it Journal of Diabetes Science and Technology}, vol. {\it 7}, no. {\it 6}, pp. 1470--1483, 2013.

\bibitem{MPC_AP5}
    M. Breton et al., ``Fully Integrated Artificial Pancreas in Type 1 Diabetes Modular Closed-Loop Glucose Control Maintains Near Normoglycemia,'' {\it Diabetes}, vol. {\it 61}, no. {\it 9}, pp. 2230--2237, 2012. 

\bibitem{NCS_AP1}
    P. Keith-Hynes et al., ``The Diabetes Assistant: A Smartphone-Based System for Real-Time Control of Blood Glucose,'' {\it Electronics}, vol. {\it 3}, no. {\it 4}, pp. 609--623, 2014.

\bibitem{NCS_AP2}
    R. A. Lal et al., ``Realizing a Closed-Loop (Artificial Pancreas) System for the Treatment of Type 1 Diabetes,'' {\it Endocrine Reviews}, vol. {\it 40}, no. {\it 6}, pp. 1521--1546, 2019.

\bibitem{MPC_and_RL}
    R. Reiter et al., ``Synthesis of Model Predictive Control and Reinforcement Learning: Survey and Classification,'' {\it Annual Reviews in Control}, vol. {\it 61}, 101045, 2026.

\bibitem{RL_book}
    R. S. Sutton and A. G. Barto, {\it Reinforcement Learning: An Introduction Second Edition}, MIT Press, 2018. 

\bibitem{DRL_book}
    H. Dong et al., {\it Deep Reinforcement Learning Fundamentals, Research and Applications}, Springer, 2021.

\bibitem{DR1}
    L. Da et al., ``A Survey of Sim-to-Real Methods in RL: Progress, Prospects and Challenges with Foundation Models,'' {\it arXiv Preprint}, arXiv:2502.13187, 2025.

\bibitem{DR2}
    W. Zhao et al., ``Sim-to-Real Transfer in Deep Reinforcement Learning for Robotics: a Survey,'' in Proc. of {\it IEEE Symposium Series on Computational Intelligence}, pp. 737--744, 2020.

\bibitem{Lee}
    S. Lee et al., ``Toward a Fully Automated Artificial Pancreas System Using a Bioinspired Reinforcement Learning Design: In Silico Validation,'' {\it IEEE Journal of Biomedical and Health Informatics}, vol. {\it 25},\break   no. {\it 2}, pp. 536--546, 2021. 

\bibitem{NCSs}
    X.-M. Zhang et al., ``Networked Control Systems: A Survey of Trends and Techniques,'' {\it IEEE/CAA Journal of Automatica Sinica}, vol. {\it 7}, no. {\it 1}, pp. 1--17, 2020.

\bibitem{etc_stc}
    W. P. M. H. Heemels et al., ``An Introduction to Event-Triggered and Self-Triggered Control,'' Proc. of {\it 2012 IEEE 51st IEEE Conference on Decision and Control (CDC)}, pp. 3270--3285, 2012. 

\bibitem{ETC_DRL_1}
    N. Funk et al., ``Learning Event-Triggered Control from Data through Joint Optimization,'' {\it IFAC Journal of Systems and Control}, vol. {\it 16}, 100144, 2021.

\bibitem{ETC_DRL_2}
    A. Termehchi and M. Rasti, ``A Learning Approach for Joint Design of Event-triggered Control and Power-Efficient Resource Allocation,'' {\it IEEE Transactions on Vehicular Technology}, vol. {\it 71}, no. {\it 6}, pp. 6322--6334, 2022.

\bibitem{ETC_DRL_3}
    L. Kesper et al., ``Toward Multi-Agent Reinforcement Learning for Distributed Event-Triggered Control,'' in Proc. of {\it 5th Annual Conference on Learning for Dynamics and Control}, vol. {\it 211}, pp. 1072--1085, 2023.

\bibitem{SMDP}
    R. S. Sutton et al., ``Between MDPs and Semi-MDPs: A Framework for Temporal Abstraction in Reinforcement Learning,'' {\it Artificial Intelligence}, vol. {\it 112}, no. {\it 1-2}, pp. 181--211, 1999.

\bibitem{fuzzy_AP}
    R. Mauseth et al., ``Use of a ``Fuzzy Logic'' Controller in a Closed-Loop Artificial Pancreas,'' {\it Diabetes Technology \& Therapeutics}, vol. {\it 15},\break   no. {\it 8}, pp. 628--633, 2013. 

\bibitem{RL_AP1}
    M. K. Bothe et al., ``The Use of Reinforcement Learning Algorithms to Meet the Challenges of an Artificial Pancreas,'' {\it Expert Review of Medical Devices}, vol. {\it 10}, no. {\it 5}, pp. 661--673, 2013.

\bibitem{RL_AP2}
    E. Daskalaki et al., ``Model-Free Machine Learning in Biomedicine: Feasibility Study in Type 1 Diabetes,'' {\it PLoS One}, vol. {\it 11}, no. {\it 7}, e0158722, 2016.

\bibitem{RL_AP3}
    Q. Sun et al., ``A Dual Mode Adaptive Basal-Bolus Advisor Based on Reinforcement Learning,'' {\it IEEE Journal of Biomedical and Health Informatics}, vol. {\it 23}, no. {\it 6}, pp. 2633--2641, 2019.

\bibitem{RL_AP4}
    M Tejedor et al., ``Reinforcement Learning Application in Diabetes Blood Glucose Control: A Systematic Review,'' {\it Artificial Intelligence In Medicine}, vol. {\it 104}, 101836, 2020.

\bibitem{RL_AP5}
    I. Fox et al., ``Deep Reinforcement Learning for Closed-Loop Blood Glucose Control,'' in Proc. of {\it Machine Learning for Healthcare Conference}, vol. {\it 126}, pp. 508--536, 2020.

\bibitem{Android_APS}
    ``AndroidAPS,'' [Online]. Available: \url{https://androidaps.readthedocs.io}

\bibitem{STC_DRL_1}
    R. Wang et al., ``Deep Reinforcement Learning for Continuous-time Self-triggered Control,'' {\it IFAC Papers Online}, vol. {\it 54}, no. {\it 14}, pp. 203--208, 2021.

\bibitem{STC_DRL_2}
    H. Wan et al., ``Model-Free Self-Triggered Control Based on Deep Reinforcement Learning for Unknown Nonlinear Systems,'' {\it International Journal of Robust and Nonlinear Control}, vol. {\it 33}, no. {\it 3}, pp. 2238--2250, 2023.

\bibitem{PG}
    R. S. Sutton et al., ``Policy Gradient Methods for Reinforcement Learning with Function Approximation,'' in Proc. of {\it Advances in Neural Information Processing Systems 12 (NIPS1999)}, pp. 1057--1063, 1999.

\bibitem{GAE}
    J. Schulman et al., ``High-Dimensional Continuous Control Using Generalized Advantage Estimation,'' {\it arXiv Preprint}, arXiv: 1506.02438, 2015.

\bibitem{A3C}
    V. Mnih et al., ``Asynchronous Methods for Deep Reinforcement Learning,'' in Proc. of {\it The 33rd International Conference on Machine Learning}, vol. {\it 48}, pp. 1928--1937, 2016.

\bibitem{TRPO}
    J. Schulman et al., ``Trust Region Policy Optimization,'' in Proc. of {\it the 32nd International Conference on Machine Learning}, vol. {\it 37}, pp. 1889--1897, 2015.

\bibitem{PPO}
    J. Schulman et al., ``Proximal Policy Optimization,'' {\it arXiv Preprint}, arXiv: 1707.06347, 2016.

\bibitem{Adam}
    D. P. Kingma and J. Ba, ``Adam: A Method for Stochastic Optimization,'' {\it arXiv Preprint}, arXiv:1412.6980, 2014.

\bibitem{PPO_1}
L. Engstrom et al., ``Implementation Matters in Deep RL: A Case Study on PPO and TRPO,'' {\it arXiv Preprint}, arXiv:2005.12729, 2020.

\bibitem{PPO_2}
M. Andrychowicz et al., ``What Matters In On-Policy Reinforcement Learning? A Large-Scale Empirical Study,'' {\it arXiv Preprint}, arXiv:2006.05990, 2020.

\bibitem{Numerical_ODE}
    E. Hairer et al., {\it Solving Ordinary Differential Equations I}, Springer, 1993.

\bibitem{Meal_Simulation_Model}
    C. D. Man et al., ``Meal Simulation Model of the Glucose-Insulin System,'' {\it IEEE Transactions on Biomedical Engineering}, vol. {\it 54}, no. {\it 10}, pp. 1740--1749, 2007.

\bibitem{TIR}
    T. Battelino et al., ``Clinical Targets for Continuous Glucose Monitoring Data Interpretation: Recommendations From the International Consensus on Time in Range,'' {\it Diabetes Care}, vol. {\it 42}, no. {\it 8}, pp. 1593--1603, 2019.

\bibitem{EGAE}
    Y. Chen et al., ``Reinforcement Learning for Robot Navigation with Adaptive Forward Simulation Time (AFST) in a Semi-Markov Model,'' Proc. of {\it IEEE/RSJ International Conference on Intelligent Robots and Systems (IROS)}, pp. 3597--3604, 2023.

\bibitem{Simglucose}
    J. Xie, ``Simglucose v0.2.1,'' [Online]. Available: \url{https://github.com/jxx123/simglucose?tab=readme-ov-file}.

\bibitem{CIRL}
    M. Bloor et al., ``Control-Informed Reinforcement Learning for Chemical Processes,'' {\it Industrial \& Engineering Chemistry Research}, vol. {\it 64}, no. {\it 9}, pp. 4966--4978, 2026.
\end{thebibliography}
\end{document}